\documentclass[amssymb,aps,prd,floatfix,twocolumn]{revtex4}
\usepackage[intlimits]{amsmath}
\usepackage{amssymb}
\usepackage{graphicx}
\usepackage{tensor}
\usepackage[dvipsnames]{xcolor}

\begin{document}

\author{Patryk Mach}\email{patryk.mach@uj.edu.pl}
\affiliation{Instytut Fizyki Teoretycznej, Uniwersytet Jagiello\'{n}ski, \L ojasiewicza 11,  30-348 Krak\'ow, Poland}
\author{Andrzej Odrzywo{\l}ek}\email{andrzej.odrzywolek@uj.edu.pl}
\affiliation{Instytut Fizyki Teoretycznej, Uniwersytet Jagiello\'{n}ski, \L ojasiewicza 11,  30-348 Krak\'ow, Poland}

\title{Accretion of the relativistic Vlasov gas onto a moving Schwarzschild black hole: Exact solutions}

\begin{abstract}
We derive an exact, axially symmetric solution representing stationary accretion of the relativistic, collisionless Vlasov gas onto a moving Schwarzschild black hole. The gas is assumed to be in thermal equilibrium at infinity, where it obeys the Maxwell-J\"{u}ttner distribution. The Vlasov equation is solved analytically in terms of suitable action-angle variables. We provide explicit expressions for the particle current density and accretion rates. In the limit of infinite asymptotic temperature of the gas, we recover the qualitative picture known from the relativistic Bondi-Hoyle-Lyttleton accretion of the perfect gas with the ultrahard equation of state, in which the mass accretion is proportional to the Lorentz factor associated with the black hole velocity. For a finite asymptotic temperature, the mass accretion rate is not in general a monotonic function of the velocity of the black hole.
\end{abstract}


\maketitle

\section{Introduction}

In this paper, we investigate stationary accretion of the relativistic Vlasov (collisionless) gas onto a moving Schwarzschild black hole. In the literature, a Newtonian counterpart of this process is known as the Bondi-Hoyle-Lyttleton accretion, a term which is used in a broad sense, referring to early works using the ballistic approximation \cite{hoyle_lyttleton, lyttleton_hoyle, bondi_hoyle}, as well as to hydrodynamical solutions \cite{bisnovatyi}. A modern review of these works can be found in \cite{edgar}.  Probably the best known general-relativistic solution describing stationary accretion of matter onto a moving black hole was derived by Petrich, Shapiro, and Teukolsky for the ultrahard equation of state \cite{petrich}. It is exceptional in the sense that for the ultrahard fluids the flow is always subsonic (the local speed of sound is equal to the speed of light), and consequently no shock waves occur in the solution. A general-relativistic version of the Bond-Hoyle-Lyttleton ballistic approximation was derived in \cite{tajeda}. In the generic, general-relativistic case, hydrodynamical solutions can be computed numerically \cite{font, font_ibanez, font_ibanez_papadopoulos, zanotti, blakely, lora, cruz, rezzolla_zanotti}. In all these cases the solution is obtained by solving equations of the general-relativistic hydrodynamics, assuming a fixed black hole background spacetime and the boundary conditions corresponding to the gas moving asymptotically with a constant speed.

In technical terms, the analysis of this paper is a sequel to the work of Rioseco and Sarbach \cite{Olivier,Olivier2}, who considered stationary, spherically symmetric accretion of the relativistic Vlasov gas onto the Schwarzschild black hole and developed an elegant Hamiltonian formalism for solving the relativistic Vlasov equation on the Schwarzschild background. The key element of this formalism is the construction of a suitable set of action-angle variables $(Q^\mu,P_\nu)$ (coordinates in the phase space) that trivialize the Vlasov equation. Consequently, the task of choosing the right distribution function (solution to the Vlasov equation) is reduced to a proper choice of the boundary conditions. In \cite{Olivier,Olivier2}, this choice corresponds to the gas which is asymptotically in thermal equilibrium and at rest---the distribution function is required to reduce asymptotically to the Maxwell-J\"{u}ttner distribution \cite{juttner1,juttner2}. As a result, the main task in the analysis consists not so much in the actual solving of the Vlasov equation, but rather in computing a set of physical variables (observables) that describe the flow: the particle current density, the particle number density, the energy density, the pressures. To this end, another set of phase-space coordinates is required, which allows us to control the region of the phase space available to the motion of the gas particles. In \cite{Olivier} and in this paper, these coordinates are denoted as $(t,r,\theta,\varphi,\varepsilon,m,\lambda,\chi)$.

In \cite{cieslik_mach} (a work coauthored by one of us) the formalism of Rioseco and Sarbach was used to compute solutions representing the accretion of the relativistic Vlasov gas onto the Reissner-Nordstr\"{o}m black hole, treated as a toy model for spinning black holes. In this paper, we abandon spherical symmetry. As usual for the relativistic analog of the Bondi-Hoyle-Lyttleton accretion problem, we consider accretion onto the Schwarzschild black hole with the asymptotic conditions corresponding to the gas in thermal equilibrium, boosted with a velocity $v$ in some direction (along the $z$ axis, say). Similarly to the spherically symmetric case treated in \cite{Olivier,Olivier2,cieslik_mach}, the gas in thermal equilibrium at the infinity is no longer in thermal equilibrium in the vicinity of the black hole. In analogy with \cite{Olivier,Olivier2,cieslik_mach}, we only consider the particles that travel from infinity and get absorbed by the black hole, and particles with sufficiently high angular momentum to get scattered to infinity. Particles on bounded trajectories are not taken into account. This is justified by the fact that scattering (collisions) between the particles is neglected, and we only consider the motion in a fixed background spacetime (there is no gravitational interaction between the particles). In comparison with \cite{Olivier,Olivier2,cieslik_mach}, the formulas describing the flow are considerably more complex, but we are still able to derive exact expressions for the particle density current, mass accretion rate, etc.

Perhaps the most interesting result of our analysis is the fact that the mass accretion rate is not, in general, a monotonic function of the black hole velocity, which is different from the situation known from \cite{petrich}. For the accretion of the perfect fluid with the ultrahard equation of state investigated in \cite{petrich}, the mass accretion rate is simply proportional to the Lorentz factor associated with black hole velocity. On the other hand, the general-relativistic ballistic approximation constructed in the spirit of Bondi, Hoyle, and Lyttleton in \cite{tajeda} yields the accretion rate with a minimum for the black hole speeds of the order of $0.8 \, c$, which agrees also with numerical results obtained for perfect fluids with the perfect-gas equation of state and with suitably adjusted values of the asymptotic speed of sound. Our results for the collisionless Vlasov gas interpolate between those two regimes, depending on the asymptotic temperature.

The assumptions that the gas is in thermal equilibrium at infinity, and that the collisions between particles can be neglected at the same time, should be understood as a statement about timescales in the accretion process. We assume that the particles interact only very weakly, but the asymptotic reservoir of the gas (the cloud of the gas) was given sufficient time to thermalize. In comparison, the timescale associated with the motion of the particles in the vicinity of the black hole moving through the gas is assumed to be short. The same assumptions were made in \cite{Olivier,cieslik_mach}. Probably the most natural application of this kind of models would be the accretion of the hypothetical dark matter particles onto black holes \cite{DM_Vlasov}.

The order of this paper is as follows. In Sec.\ \ref{sec:vlasovinschw} we recall the Hamiltonian approach to the Vlasov equation in the Schwarzschild spacetime and introduce suitable action-angle coordinates. In Sec.\ \ref{sec:boostedmaxwell} we discuss the boosted Maxwell-J\"{u}ttner distribution, serving as an asymptotic condition for our accretion model. We derive the expression for the asymptotic distribution function, expressed both in terms of the spherical phase-space coordinates and in the action-angle variables. In Sec.\ \ref{sec:momentumintegrals} we compute several quantities expressed as suitable integrals over momenta, in particular the particle current density and the mass accretion rate. Numerical results are discussed in Sec.\ \ref{sec:numeric}. Section \ref{sec:conclusions} contains a few concluding remarks.

In this paper we use geometric units with $c = G = 1$, where $c$ denotes the speed of light, and $G$ is the gravitational constant. Spacetime dimensions are labeled with Greek indices $\mu = 0, 1, 2, 3$. Spatial dimensions are denoted with Latin indices $i = 1, 2, 3$. The signature of the metric is $(-,+,+,+)$.

\section{Vlasov equation in the Schwarzschild spacetime}
\label{sec:vlasovinschw}

\subsection{Hamiltonian description of the geodesic motion}

We start this section with recalling the basics of the Hamiltonian description of the geodesic motion and the Vlasov equation (in the Hamiltonian framework).

Let $g_{\mu\nu}$ denote the components of the spacetime metric, and let $(x^\mu)$ be local coordinates. We consider timelike geodesics $\Gamma$ parametrized with a parameter $\tau$. The momenta $p^\mu$ are defined as four-vectors tangent to $\Gamma$: $p^\mu = d x^\mu/d\tau$. The Hamiltonian of a free particle can be chosen as
\[ H(x^\mu,p_\nu) = \frac{1}{2}g^{\mu \nu}(x^\alpha) p_\mu p_\nu, \]
where $(x^\mu,p_\nu)$ are treated as canonical variables. Note that we deliberately choose to work with contravariant representation of the coordinates $x^\mu$ and covariant momenta $p_\mu$ (one forms). It is easy to show that the equations of motion
\[ \frac{d x^\mu}{d \tau} = \frac{\partial H}{\partial p_\mu}, \quad \frac{d p_\nu}{d \tau} = - \frac{\partial H}{\partial x^\nu} \]
imply standard geodesic equations
\[ \frac{d^2 x^\mu}{d \tau^2} + \Gamma^\mu_{\alpha \beta} \frac{d x^\alpha}{d \tau} \frac{d x^\beta}{d \tau} = 0, \]
where $\Gamma^\mu_{\alpha \beta}$ denotes the Christoffel symbols associated with the metric $g$. Here, a somewhat subtle point is the choice of the parameter $\tau$. We require that $H = \frac{1}{2} g^{\mu \nu} p_\mu p_\nu = - \frac{1}{2}m^2$, where $m$ denotes the rest mass of the particle. As a consequence, $\tau = s/m$, where $s$ is the proper time. The four-velocity $u^\mu = d x^\mu/ds$ is normalized as $g^{\mu \nu}u_\mu u_\nu = -1$.

\subsection{Vlasov equation}

The Vlasov gas of noncolliding particles is described by the probability function $f = f(x^\mu,p_\nu)$, which should be invariant along a geodesic, i.e., 
\begin{eqnarray*}
\lefteqn{\frac{d}{d \tau} f(x^\mu(\tau),p_\nu(\tau))  =  \frac{dx^\mu}{d \tau} \frac{\partial f}{\partial x^\mu} + \frac{d p_\nu}{d \tau} \frac{\partial f}{\partial p_\nu} } \\
&& = \frac{\partial H}{\partial p_\mu} \frac{\partial f}{\partial x^\mu} - \frac{\partial H}{\partial x^\nu} \frac{\partial f}{\partial p_\nu} = \{ H, f \} = 0,
\end{eqnarray*}
where $\{ \cdot, \cdot\}$ denotes the Poisson bracket. In more explicit terms, the Vlasov equation
\begin{equation}
\label{vlasoveq}
\frac{\partial H}{\partial p_\mu} \frac{\partial f}{\partial x^\mu} - \frac{\partial H}{\partial x^\nu} \frac{\partial f}{\partial p_\nu} = 0
\end{equation}
can be written as \cite{andreasson,rendall}
\[ g^{\mu\nu} p_\nu \frac{\partial f}{\partial x^\mu} - \frac{1}{2} p_\alpha p_\beta \frac{\partial g^{\alpha \beta}}{\partial x^\mu} \frac{\partial f}{\partial p_\mu} = 0. \]

An important ingredient of the relativistic Vlasov model is the way of associating the energy momentum tensor $T_{\mu \nu}$ to a given distribution function. It is assumed that $T_{\mu \nu}$ can be expressed as
\begin{equation}
\label{tmunu}
T_{\mu \nu}(x) = \int p_\mu p_\nu f(x,p) \mathrm{dvol}_x(p),
\end{equation}
where
\[ \mathrm{dvol}_x(p) = \sqrt{- \mathrm{det}[g^{\mu\nu}(x)]} dp_0dp_1dp_2dp_3. \]
At the same time the particle current density is given by
\begin{equation}
\label{jgeneral}
J_\mu = \int p_\mu f (x,p) \mathrm{dvol}_x(p).
\end{equation}
The Vlasov equation can be used to show that
\begin{equation}
\label{conservationj}
\nabla_\mu J^\mu = 0
\end{equation}
and
\begin{equation}
\label{conservationtmunu}
\nabla_\mu T^{\mu \nu} = 0,
\end{equation}
where $\nabla_\mu$ denotes the covariant derivative associated with the metric $g$ \cite{rezzolla_zanotti}. The particle number density can be defined covariantly as
\begin{equation}
\label{ngeneral}
n = \sqrt{-J_\mu J^\mu}.
\end{equation}

\subsection{Horizon-penetrating coordinates}

In this work we consider a gas of free particles on the fixed Schwarzschild background spacetime with the metric
\[ g = - N(\bar r) d\bar t^2 + \frac{1}{N(\bar r)} d \bar r^2 + \bar r^2 (d \theta^2 + \sin^2 \theta d \varphi^2), \]
where $N = 1 - 2M/\bar r$. As usual for accretion problems, it is convenient to work in horizon-penetrating coordinates. We introduce a standard transformation $(\bar t, \bar r) \to (t,r)$ given by
\[ t = \bar t + \int^{\bar r} \left[ \frac{1}{N(s)} - \eta(s) \right] ds,  \quad r = \bar r, \]
where $\eta$ is a function, which effectively controls the time foliation. In terms of coordinates $(t,r)$ the metric can be expressed as
\begin{eqnarray}
g & = & -N dt^2 + 2(1 - N \eta) dt dr + \eta (2 - N \eta) dr^2 \nonumber \\
&& + r^2 (d \theta^2 + \sin^2 \theta d \varphi^2).
\label{efgeneral}
\end{eqnarray}
The contravariant metric components can be easily computed:
\[ g^{tt} = \eta (-2 + N \eta), \quad g^{tr} = 1 - N \eta, \quad g^{rr} = N. \]
Note that
\begin{equation}
\label{metricunity}
\left(g^{tr}\right)^2 - g^{rr}g^{tt} = 1.
\end{equation}
The function $\eta$ defines the time foliation. A popular choice (sometimes referred to as the Eddington-Finkelstein coordinates) is to set $\eta \equiv 1$. We adhere to this choice in the discussion in Sec.\ \ref{sec:boostedmaxwell} and in our numerical examples.

\subsection{Action-angle variables in the Schwarzschild spacetime}
\label{sectionactionangle}

The Hamiltonian of a free particle in the Schwarzschild spacetime endowed with metric (\ref{efgeneral}) reads
\begin{eqnarray}
H & = & \frac{1}{2} \left[ g^{tt}(r) p_t^2 + 2 g^{tr}(r) p_t p_r + g^{rr}(r) p_r^2 \right. \nonumber \\
& & \left. + \frac{1}{r^2} \left( p_\theta^2 + \frac{p_\varphi^2}{\sin^2 \theta} \right) \right].
\label{hamiltonian_schw}
\end{eqnarray}
This form implies the following conserved quantities: $E = -p_t$ (because $H$ does not depend on $t$), $l_z = p_\varphi$ (since $H$ does not depend on $\varphi$), and $m = \sqrt{-2H}$ (because $H$ does not depend explicitly on the parameter $\tau$). An elementary calculation shows that the total angular momentum
\begin{equation}
\label{ang_mom}
l = \sqrt{p_\theta^2 + \frac{p_\varphi^2}{\sin^2 \theta}}
\end{equation}
is also conserved.

For a geodesic with fixed $E$, $l_z$, $l$, and $m$, the radial momentum can be computed as
\begin{eqnarray}
p_r & = & \frac{g^{tr} E + \epsilon_r \sqrt{\left[ (g^{tr})^2 - g^{tt} g^{rr} \right] E^2 - g^{rr} \left( m^2 + \frac{l^2}{r^2} \right)}}{g^{rr}} \nonumber \\
& = & \frac{(1 - N \eta) E + \epsilon_r \sqrt{E^2 - \tilde U_{l,m}(r)}}{N},
\label{prformula}
\end{eqnarray}
where we have used Eq.\ (\ref{metricunity}) and introduced the effective potential
\[ \tilde U_{l,m}(r) = N \left( m^2 + \frac{l^2}{r^2} \right). \]
We also denote $\epsilon_r = \pm 1$, distinguishing between ingoing and outgoing particles. The $p_\theta$ component is given by
\begin{equation}
\label{pthetaformula}
p_\theta = \epsilon_\theta \sqrt{l^2 - \frac{l_z^2}{\sin^2 \theta}}, \end{equation}
where, similarly to $\epsilon_r$, $\epsilon_\theta = \pm 1$.

Consider a motion along a geodesic $\Gamma$ with constant $E, l_z, l$, and $m$. Define the so-called abbreviated action \cite{Hand,Goldstein}
\[ S = \int_\Gamma p_\mu dx^\mu = -E t + l_z \varphi + \int_\Gamma p_r dr + \int_\Gamma p_\theta d\theta. \]
Here $p_r$ and $p_\theta$ are given by Eqs.\ (\ref{prformula}) and (\ref{pthetaformula}), respectively. We now introduce the following canonical transformation: $(t,r,\theta,\varphi,p_t,p_r,p_\theta,p_\varphi) \to (Q^\mu,P_\nu)$, where the new momenta are constant:
\begin{widetext}
\begin{eqnarray*}
P_0 & = & m = \sqrt{ -g^{tt}(r)(p_t)^2 - 2 g^{tr}(r) p_t p_r  - g^{rr}(r) (p_r)^2 - \frac{1}{r^2} \left( p_\theta^2 + \frac{p_\varphi^2}{\sin^2 \theta} \right)}, \\
P_1 & = & E = - p_t, \\
P_2 & = & l_z = p_\varphi, \\
P_3 & = & l = \sqrt{p_\theta^2 + \frac{p_\varphi^2}{\sin^2 \theta}},
\end{eqnarray*}
\end{widetext}
and the corresponding conjugate variables are defined as
\begin{subequations}
\begin{eqnarray}
\label{q0}
Q^0 = \frac{\partial S}{\partial m} & = & - m \int_\Gamma \frac{dr}{-g^{tr} E + g^{rr} p_r}, \\
\label{q1}
Q^1 = \frac{\partial S}{\partial E} & = & - t + \int_\Gamma \frac{g^{tt} E - g^{tr} p_r}{g^{tr} E - g^{rr} p_r} dr, \\
\label{q2}
Q^2 = \frac{\partial S}{\partial l_z} & = & \varphi - l_z \int_\Gamma \frac{d \theta}{p_\theta \sin^2 \theta}, \\
Q^3 = \frac{\partial S}{\partial l} & = & - l \int_\Gamma \frac{dr}{r^2 \left( -g^{tr} E + g^{rr} p_r \right)} \nonumber \\
&& + l \int_\Gamma \frac{d \theta}{p_\theta},
\label{q3}
\end{eqnarray}
\end{subequations}
where again all integrals are understood as line integrals along geodesics with constant $m$, $E$, $l_z$, and $l$.

The main advantage of the new coordinates $(P_\mu, Q^\nu)$ is that they trivialize the Vlasov equation. The Hamiltonian reads simply $H = -P_0^2/2$. Since the Poisson bracket is covariant with respect to canonical transformations, we have
\begin{eqnarray*}
\lefteqn{\frac{\partial H}{\partial p_\mu} \frac{\partial}{\partial x^\mu} - \frac{\partial H}{\partial x^\nu} \frac{\partial}{\partial p_\nu}} \\
&& = \frac{\partial H}{\partial P_\mu} \frac{\partial}{\partial Q^\mu} - \frac{\partial H}{\partial Q^\nu} \frac{\partial}{\partial P_\nu} = - P_0 \frac{\partial}{\partial Q^0}.
\end{eqnarray*}
Consequently, the Vlasov equation takes the form
\begin{equation}
\label{vlasovQ0}
\frac{\partial f}{\partial Q^0} = 0,
\end{equation}
and its general solution can be locally written as
\begin{equation}
\label{general_solution}
f(x^\mu, p_\nu) = \mathcal F (Q^1, Q^2, Q^3, P_0, P_1, P_2, P_3).
\end{equation}

Further restrictions on the form of $f$ follow from the symmetry assumptions. This can be understood by computing the lifts of the Killing vectors generating the symmetries of the spacetime to the cotangent bundle with local coordinates $(x^\mu,p_\nu)$. Given a Killing vector
\[ \xi_x = \xi^\mu(x) \left. \frac{\partial}{\partial x^\mu} \right|_{x}, \]
we compute its lift as
\[ \hat \xi_{(x,p)} = \left. \xi^\mu (x) \frac{\partial}{\partial x^\mu}\right|_{(x,p)} - \left. p_\alpha \frac{\partial \xi^\alpha}{\partial x^\mu}(x) \frac{\partial}{\partial p_\mu} \right|_{(x,p)}. \]
The case of stationary and axially symmetric flows is relatively simple, as we are only interested in two Killing vectors: $k = \partial/\partial t$ and $\xi_3 = \partial/\partial \varphi$, with the corresponding lifts also given by
\[ \hat k = \frac{\partial}{\partial t}, \quad \hat \xi_3 = \frac{\partial}{\partial \varphi}. \]
Expressing $\hat k$ and $\hat \xi_3$ in terms of the coordinates $(Q^\mu,P_\nu)$ we get
\[ \hat k = \frac{\partial}{\partial t} = \frac{\partial Q^\mu}{\partial t} \frac{\partial}{\partial Q^\mu} + \frac{\partial P_\nu}{\partial t} \frac{\partial}{\partial P_\nu} = - \frac{\partial}{\partial Q^1}. \]
\[ \hat \xi_3 = \frac{\partial}{\partial \varphi} = \frac{\partial Q^\mu}{\partial \varphi} \frac{\partial}{\partial Q^\mu} + \frac{\partial P_\nu}{\partial \varphi} \frac{\partial}{\partial P_\nu} = \frac{\partial}{\partial Q^2}. \]
Hence, a stationary distribution should be locally independent of $Q^1$, and an axially symmetric one should be locally independent of $Q^2$. The above reasoning is closely related with the so-called Jeans theorem proved for Newtonian Vlasov-Poisson systems in \cite{batt}. It states that every spherically symmetric, stationary solution of the Vlasov-Poisson system can be globally expressed as a function of the energy and the angular momentum only. The general relativistic counterpart of the Jeans theorem is known to be false---counterexamples were constructed in \cite{schaeffer}. Since in this paper we are dealing with the fixed Schwarzschild background, and we will restrict ourselves to a very special subset of trajectories relevant to the accretion process (excluding the particles on bounded orbits), we will search for a global solution of the form
\[ f(x^\mu, p_\nu) = \mathcal F (Q^3, P_0, P_1, P_2, P_3). \]

In the following, it will be convenient to use dimensionless quantities, introduced in \cite{Olivier}. We define
\begin{eqnarray*}
t & = & M \tilde \tau, \\
r & = & M \xi, \\
p_r & = & m \pi_\xi, \\
p_\theta & = & M m \pi_\theta, \\
E & = & m \varepsilon, \\
l & = & M m \lambda, \\
l_z & = & M m \lambda_z. 
\end{eqnarray*}
Note that
\[ N = 1 - \frac{2}{\xi}, \]
and with $\eta \equiv 1$,
\[ g^{tt} = - \left( 1 + \frac{2}{\xi} \right), \quad g^{tr} = \frac{2}{\xi}, \quad g^{rr} = 1 - \frac{2}{\xi}. \]
In terms of dimensionless quantities Eq.\ (\ref{pthetaformula}) takes the form
\[ \pi_\theta = \epsilon_\theta \sqrt{\lambda^2 - \frac{\lambda_z^2}{\sin^2 \theta}}, \]
while for $\pi_\xi$ we get
\begin{equation}
\label{pixi}
\pi_\xi = \frac{(1 - N \eta) \varepsilon + \epsilon_r \sqrt{\varepsilon^2 - U_\lambda(\xi)}}{N},
 \end{equation}
where the dimensionless effective potential reads
\[ U_\lambda(\xi) = N \left(1 + \frac{\lambda^2}{\xi^2} \right) = \left(1 - \frac{2}{\xi} \right) \left(1 + \frac{\lambda^2}{\xi^2} \right).  \]

For $Q^3$ we obtain
\begin{eqnarray}
\label{Q3}
Q^3 & = & - \lambda \int_\Gamma \frac{d \xi}{\xi^2 \left( - g^{tr} \varepsilon + g^{rr} \pi_\xi \right)} + \lambda \int_\Gamma \frac{d \theta}{\pi_\theta} \\
& = & \epsilon_r X(\xi,\varepsilon,\lambda) - \epsilon_r \frac{\pi}{2} - \epsilon_\theta \arctan \left( \frac{\lambda \cot \theta}{\sqrt{\lambda^2 - \frac{\lambda_z^2}{\sin^2 \theta}}} \right), \nonumber
\end{eqnarray}
where
\begin{eqnarray}
X(\xi,\varepsilon,\lambda) = \lambda \int_\xi^\infty \frac{d \xi^\prime}{{\xi^\prime}^2 \sqrt{\varepsilon^2 - \left(1 - \frac{2}{\xi^\prime} \right) \left( 1 + \frac{\lambda^2}{{\xi^\prime}^2} \right)}}.
\label{X}
\end{eqnarray}
For future convenience, we chose the reference point in computing the integrals in $Q^3$ so that for $\xi \to \infty$, $X(\xi,\varepsilon,\lambda) \to 0$. The above convention is consistent with (and in fact motivated by) the choice made in Sec.\ \ref{sec:boostedmaxwell} for the flat spacetime. The integral in Eq.\ (\ref{X}) can be computed analytically and expressed for example in terms of the inverse of the Weierstrass elliptic function $\wp$ (or rather its suitable restrictions). We postpone the derivation of this result to Appendix \ref{appendixa}.

In deriving Eq.\ (\ref{Q3}) we have assumed that the signs $\epsilon_r$ and $\epsilon_\theta$ remain constant along a given trajectory. Since both signs can change along a single trajectory, in our analysis we will divide the trajectories into segments characterized by constant $\epsilon_r$ and $\epsilon_\theta$. For example, the contribution to the particle current density due to particles scattered by the black hole will be computed by considering separately segments of the orbits corresponding to incoming and outcoming particles.

\section{Boosted Maxwell-J\"{u}ttner distribution}
\label{sec:boostedmaxwell}

\subsection{Maxwell-J\"{u}ttner distribution in the Minkowski spacetime}

In the light of the formalism described in the preceding sections, the main task in the search for solutions representing the accretion onto a moving Schwarzschild black hole consists in choosing the appropriate distribution function $f = \mathcal F(Q^3,P_0,P_1,P_2,P_3)$. We will do this by demanding that it should correspond asymptotically to a boosted Maxwell-J\"{u}ttner distribution.

We start by considering the Maxwell-J\"{u}ttner distribution in the Minkowski spacetime. The metric is written as
\begin{eqnarray*}
g & = & -dt^2 + dx^2 + dy^2 + dz^2 \\
& = & -dt^2 + dr^2 + r^2 (d \theta^2 + \sin^2 \theta d \varphi^2),
\end{eqnarray*}
either in Cartesian $(t,x,y,z)$ or spherical $(t,r,\theta,\varphi)$ coordinates. The timelike Killing vector reads $k^\mu = (1,0,0,0)$ (in both coordinate systems). We write the Maxwell-J\"{u}ttner distribution for the particles of the same mass $m$ (the so-called simple gas) as
\begin{equation}
\label{juttnera}
f(x^\mu,p_\nu) = \delta\left( \sqrt{-p_\mu p^\mu} - m \right) F(x^\mu,p_\nu),
\end{equation}
where
\begin{equation}
\label{juttnerb}
F(x^\mu,p_\nu) = \alpha \exp \left(\frac{\beta}{m} k^\mu p_\mu \right) = \alpha \exp \left( \frac{\beta p_t}{m} \right),
\end{equation}
and $\alpha$ and $\beta$ are constants. It is obviously a Lorentz scalar, as it should be. Note that the Dirac delta term in Eq.\ (\ref{juttnera}) can be omitted in the formalism in which the momenta are a priori required to satisfy the mass shell condition. In the following, we will refer to both $f$ and $F$ given by Eqs.\ (\ref{juttnera}) and (\ref{juttnerb}) as to Maxwell-J\"{u}ttner distribution functions. The constant $\beta$ is related to the temperature $T$ as
\[ \beta = \frac{m}{k_\mathrm{B} T}, \]
where $k_\mathrm{B}$ denotes the Boltzmann constant. The constant $\alpha$ can be related to the particle number density
\begin{equation}
\label{ninfty}
n_\infty = 4 \pi \alpha m^4 \frac{K_2(\beta)}{\beta},
\end{equation}
where $K_2$ is the modified Bessel function of the second kind \cite{israel}. We use the subscript $\infty$ in Eq.\ (\ref{ninfty}), since $n_\infty$ will be used in subsequent sections to denote the asymptotic value of the particle number density. Note that the convention concerning the constant $\beta$ is changed here with respect to \cite{Olivier,cieslik_mach}. In turn, we adhere to the original notation used by J\"{u}ttner \cite{juttner1} and also Israel in \cite{israel}.

Boosting this distribution with the velocity $v$ along the $z$ axis is simple. Under a Lorentz boost, $k^\mu$ transforms as
\[ {k^\prime}^\mu = ({k^\prime}^t,{k^\prime}^x,{k^\prime}^y,{k^\prime}^z)= (\gamma, 0, 0, -\gamma v), \quad \gamma = \frac{1}{\sqrt{1 - v^2}} \]
in Cartesian coordinates. Thus, the boosted Maxwell-J\"{u}ttner distribution reads
\[ f^\prime (x^\mu,p_\nu) = \delta\left( \sqrt{-p_\mu p^\mu} - m \right) F^\prime (x^\mu,p_\nu), \]
where
\[ F^\prime (x^\mu,p_\nu) = \alpha \exp \left(\frac{\beta}{m} {k^\prime}^\nu p_\nu \right) = \alpha \exp \left[ \frac{\beta}{m} \gamma (p_t - v p_z) \right]. \]
Note that we boost the gas in the negative direction of the $z$ axis---this will correspond to the black hole moving in the positive direction. In terms of spherical coordinates $(t,r,\theta,\varphi)$ related to the Cartesian ones by
\begin{eqnarray*}
x & = & r \cos \varphi \sin \theta, \\
y & = & r \sin \varphi \sin \theta, \\
z & = & r \cos \theta,
\end{eqnarray*}
we get
\begin{equation}
\label{boostedf}
F^\prime (x^\mu,p_\nu) = \alpha \exp \left\{ \frac{\beta}{m} \gamma \left[ p_t - v \left( \cos \theta p_r - \frac{\sin \theta}{r} p_\theta \right) \right] \right\}.
\end{equation}
For simplicity, we will omit the prime in $f$ and $F$ in the remainder of this paper.

As a consistency test one can check that the above distribution function satisfies the Vlasov equation (\ref{vlasoveq}). In terms of the spherical coordinates, the Hamiltonian function of a free particle in the Minkowski spacetime can be written as
\begin{equation}
\label{hamiltonian_mink}
H = \frac{1}{2} \left[ - p_t^2 + p_r^2 + \frac{1}{r^2} \left( p_\theta^2 + \frac{p_\varphi^2}{\sin^2 \theta} \right) \right].
\end{equation}
Showing that Eq.\ (\ref{vlasoveq}) is satisfied with $H$ and $f$ (or $F$) given, respectively, by Eqs.\ (\ref{hamiltonian_mink}) and (\ref{boostedf}) is a straightforward calculation.

\subsection{Boosted Maxwell-J\"{u}ttner distribution and action-angle variables}
\label{boostedmaxwell}

Similarly to the Schwarzschild case, $E = - p_t$, $l_z = p_\varphi$, $m = \sqrt{-2 H}$, and $l$ given by Eq.\ (\ref{ang_mom}) are also constants of motion governed by the Hamiltonian (\ref{hamiltonian_mink}).

For a geodesic parametrized by fixed $E$, $l_z$,  $l$, and $m$, the momentum component $p_\theta$ is still given by Eq.\ (\ref{pthetaformula}), but the formula for $p_r$ is much simpler, namely
\[ p_r = \epsilon_r \sqrt{E^2 - m^2 - \frac{l^2}{r^2}}, \]
with $\epsilon_r = \pm 1$. This allows us to express the abbreviated action as
\begin{widetext}
\begin{eqnarray*}
S & = & \int_\Gamma p_\mu dx^\mu = -E t + l_z \varphi + \int_\Gamma p_r dr + \int_\Gamma p_\theta d\theta \\
& = & -E t + l_z \varphi + \epsilon_r \int \sqrt{E^2 - m^2 - \frac{l^2}{r^2}} dr + \epsilon_\theta \int \sqrt{l^2 - \frac{l_z^2}{\sin^2 \theta}} d\theta.
\end{eqnarray*}
Here, as before, we assume that the signs $\epsilon_r$ and $\epsilon_\theta$ are constant. The last two integrals can be computed analytically. We get, in total,
\begin{eqnarray}
S & = & -E t + l_z \varphi + \epsilon_r \left[ r \sqrt{E^2 - m^2 - \frac{l^2}{r^2}} - l \arctan \left( \frac{r}{l} \sqrt{E^2 - m^2 - \frac{l^2}{r^2}} \right) \right] \nonumber \\
&& + \epsilon_\theta \left[ - l \arctan \left( \frac{l \cot \theta}{\sqrt{l^2 - \frac{l_z^2}{\sin^2 \theta}}} \right) + l_z \arctan \left( \frac{l_z \cot \theta}{\sqrt{l^2 - \frac{l_z^2}{\sin^2 \theta}}} \right) \right] + \mathrm{const}.
\label{Smink}
\end{eqnarray}

In analogy to the Schwarzschild case, we introduce the following canonical transformation: $(t,r,\theta,\varphi,p_t,p_r,p_\theta,p_\varphi) \to (Q^\mu,P_\nu)$, where the new momenta are constant:
\begin{subequations}
\label{pmink}
\begin{eqnarray}
\label{p0mink}
P_0 & = & m = \sqrt{ (p_t)^2  - (p_r)^2 - \frac{1}{r^2} \left( p_\theta^2 + \frac{p_\varphi^2}{\sin^2 \theta} \right)}, \\
\label{p1mink}
P_1 & = & E = - p_t, \\
\label{p2mink}
P_2 & = & l_z = p_\varphi, \\
\label{p3mink}
P_3 & = & l = \sqrt{p_\theta^2 + \frac{p_\varphi^2}{\sin^2 \theta}},
\end{eqnarray}
\end{subequations}
and the corresponding conjugate variables are defined as
\begin{subequations}
\label{qmink}
\begin{eqnarray}
\label{q0mink}
Q^0 = \frac{\partial S}{\partial m} & = & - \frac{m r \epsilon_r \sqrt{E^2 - m^2 - \frac{l^2}{r^2}}}{E^2 - m^2} = - \frac{r p_r}{(p_r)^2 + \frac{l^2}{r^2}} \sqrt{(p_t)^2 - (p_r)^2 - \frac{l^2}{r^2}}, \\
\label{q1mink}
Q^1 = \frac{\partial S}{\partial E} & = & - t + \frac{E r \epsilon_r \sqrt{E^2 - m^2 - \frac{l^2}{r^2}}}{E^2 - m^2} = - t - \frac{r p_t p_r}{(p_r)^2 + \frac{l^2}{r^2}}, \\
\label{q2mink}
Q^2 = \frac{\partial S}{\partial l_z} & = & \varphi + \arctan \left( \frac{l_z \cot \theta}{\sqrt{l^2 - \frac{l_z^2}{\sin^2 \theta}}} \right) = \varphi + \arctan \left( \frac{p_\varphi \cot \theta}{p_\theta} \right), \\
Q^3 = \frac{\partial S}{\partial l} & = & - \epsilon_r \arctan \left( \frac{r}{l} \sqrt{E^2 - m^2 - \frac{l^2}{r^2}} \right) - \epsilon_\theta \arctan \left( \frac{l \cot \theta}{\sqrt{l^2 - \frac{l_z^2}{\sin^2 \theta}}} \right) \nonumber \\
& = & - \arctan \left( \frac{r p_r}{l} \right) - \arctan \left( \frac{l \cot \theta}{p_\theta} \right).
\label{q3mink}
\end{eqnarray}
\end{subequations}
\end{widetext}
In the above formulas the last form is given in terms of the momenta $(p_t,p_r,p_\theta,p_\varphi)$. The total angular momentum $l$ is given by Eq.\ (\ref{ang_mom}). In deriving expressions (\ref{qmink}) we have assumed that the constant in Eq.\ (\ref{Smink}) is independent of $m$, $E$, $l_z$, and $l$. Our normalization assumed in Eq.\ (\ref{Q3}) for the coordinate $Q^3$ in the Schwarzschild metric was chosen precisely to match the normalization of Eq.\ (\ref{q3mink}).

Inverting the transformation $(t,r,\theta,\varphi,p_t,p_r,p_\theta,p_\varphi) \to (Q^\mu,P_\nu)$ is tedious. We get
\[ r p_r = - \frac{(P_1^2 - P_0^2) Q^0}{P_0} \]
and
\[ r^2 = \frac{(P_1^2 - P_0^2)(Q^0)^2}{P_0^2} + \frac{P_3^2}{P_1^2 - P_0^2}. \]
The formulas for $\theta$ and $p_\theta$ can be derived from
\[ P_3^2 = p_\theta^2 + \frac{P_2^2}{\sin^2 \theta} \]
and
\[ \frac{P_3 \cot \theta}{p_\theta} = \tan \left\{ - Q^3 - \arctan \left[ \frac{(P_0^2 - P_1^2)Q^0}{P_0 P_3} \right] \right\}. \]
We get, for instance,
\[ \sin^2 \theta = \frac{P_3^2 + P_2^2 V^2}{P_3^2 (1 + V^2)}, \]
where
\[ V = \tan \left\{ - Q^3 - \arctan \left[ \frac{(P_0^2 - P_1^2)Q^0}{P_0 P_3} \right] \right\}. \]

As the second consistency test we check that $F$ given by Eq.\ (\ref{boostedf}) does not depend on $Q^0$, as required by Eq.\ (\ref{vlasovQ0}). In principle this should be a straightforward task. The mapping $(t,r,\theta,\varphi,p_t,p_r,p_\theta,p_\varphi) \to (Q^\mu,P_\nu)$ can be inverted, and one can express $F$ in terms of new variables $(Q^\mu,P_\nu)$. However, the resulting formula is long, and it is difficult to show that the terms containing $Q^0$ cancel out. Instead, we write
\begin{eqnarray}
\label{Q0check}
\frac{\partial F}{\partial Q^0} & = & \frac{\partial x^\mu}{\partial Q^0} \frac{\partial F}{\partial x^\mu} +  \frac{\partial p_\nu}{\partial Q^0} \frac{\partial F}{\partial p_\nu} \\
& = & \frac{\partial r}{\partial Q^0} \frac{\partial F}{\partial r} + \frac{\partial \theta}{\partial Q^0} \frac{\partial F}{\partial \theta} + \frac{\partial p_r}{\partial Q^0} \frac{\partial F}{\partial p_r} + \frac{\partial p_\theta}{\partial Q^0} \frac{\partial F}{\partial p_\theta}, \nonumber
\end{eqnarray}
where on the right-hand side we only keep nonzero terms. By inverting the Jacobian $\partial (Q^\mu,P_\mu)/\partial (x^\nu,p_\nu)$ one can show that the right-hand side of Eq.\ (\ref{Q0check}) vanishes.

\subsection{Asymptotic relations}
\label{sectionasymptotic}

In fact, it suffices to consider asymptotic relations only. For $r \to \infty$, Eq.\ (\ref{boostedf}) tends to
\begin{equation}
\label{fasymptotic}
F (\xi^\mu,p_\nu) = \alpha \exp \left[ \frac{\beta}{m} \gamma \left( p_t - v \cos \theta p_r \right) \right],
\end{equation}
where the term proportional to $p_\theta$ drops out, since we only consider trajectories with finite $l$ and $l_z$, and $p_\theta$ is given by Eq.\ (\ref{pthetaformula}). On the other hand, asymptotically,
\begin{eqnarray*}
Q^3 & = & - \epsilon_r \frac{\pi}{2} - \epsilon_\theta \arctan \left( \frac{l \cot \theta}{\sqrt{l^2 - \frac{l_z^2}{\sin^2 \theta}}} \right) \\
& = & - \epsilon_r \frac{\pi}{2} - \arctan \left( \frac{l \cot \theta}{p_\theta} \right).
\end{eqnarray*}
Note that $-Q^3 - \epsilon_r \pi/2 \in (-\pi/2, \pi/2)$.
A straightforward calculation yields
\[ \cos^2 \theta = \frac{P_3^2 - P_2^2}{P_3^2} \sin^2 \left( -Q^3 - \epsilon_r \frac{\pi}{2} \right), \]
and
\[ \mathrm{sign} (\cos \theta) = \epsilon_\theta \mathrm{sign} \left[ \tan \left( -Q^3 - \epsilon_r \frac{\pi}{2} \right) \right]. \]
This gives
\[ \cos \theta = \epsilon_\theta \frac{\sqrt{P_3^2 - P_2^2}}{P_3} \sin \left( -Q^3 - \epsilon_r \frac{\pi}{2} \right). \]
Since asymptotically,
\[ p_r = \epsilon_r \sqrt{E^2 - m^2} = \epsilon_r \sqrt{P_1^2 - P_0^2},  \]
we get
\begin{widetext}
\begin{equation}   
\label{factionangle}
F = \alpha \exp \left\{ \frac{\beta}{P_0} \gamma \left[ -P_1 - v \epsilon_r \epsilon_\theta \sqrt{P_1^2 - P_0^2} \frac{\sqrt{P_3^2 - P_2^2}}{P_3} \sin \left( -Q^3 - \epsilon_r \frac{\pi}{2} \right) \right]\right\}.
\end{equation}
\end{widetext}
Let us stress that the above formula, derived based on the asymptotic relations, is actually valid everywhere in the Minkowski spacetime, provided that $p_r \neq 0$. One can substitute in Eq.\ (\ref{factionangle}) the expressions for $P_0$, $P_1$, $P_2$, $P_3$, and $Q^3$ derived in Sec.\ \ref{boostedmaxwell}, i.e., Eqs.\ (\ref{pmink}) and (\ref{q3mink}), and after some algebra obtain Eq.\ (\ref{boostedf}). This means, in particular, that Eq.\ (\ref{factionangle}) can be derived without referring to the asymptotic form given by Eq.\ (\ref{fasymptotic}) and without the assumptions on the falloff of the term $p_\theta/r$ made in deriving Eq.\ (\ref{fasymptotic}).

Standard spherical coordinates used in this section correspond to the limit $M \to 0$ in the Schwarzschild metric (\ref{efgeneral}), but also to the choice $\eta \to 1$. On the other hand, the expressions for $P_\mu$ and $Q^3$ in (\ref{factionangle}) are, in fact, independent of the choice of $\eta$. This allows us to restore the explicit dependence of some of the quantities on an arbitrary function $\eta$ in the remaining sections.

\section{Momentum integrals}
\label{sec:momentumintegrals}

\subsection{Properties of the effective potential, classification of trajectories}

The probability function given by Eq.\ (\ref{factionangle}) with the action-angle variables $(Q^\mu, P_\nu)$ computed for the Schwarzschild metric in Sec.\ \ref{sectionactionangle} constitutes the desired solution of the Vlasov equation describing the accretion onto a moving Schwarzschild black hole. However, in order to compute momentum integrals appearing in Eq.\ (\ref{jgeneral}), one needs a set of variables allowing one to control the region in the phase space available for the motion of the particles. We define such coordinates in the following subsection. In this subsection, we classify the trajectories of the particles that can reach to infinity and recall the relevant properties of the effective potential $U_\lambda(\xi)$. A detailed discussion of this topic can be found in \cite{Olivier}.

For $\lambda^2 \le 12$, the effective potential is an increasing function of $\xi$, growing from $U_\lambda = 0$ at $\xi = 2$ (the horizon) to $U_\lambda = 1$ at $\xi \to \infty$. If $\lambda^2 > 12$, there is a local maximum at
\[ \xi_\mathrm{max} = \frac{\lambda^2}{2} \left( 1 - \sqrt{1 - \frac{12}{\lambda^2}} \right), \]
and a local minimum at
\[ \xi_\mathrm{max} = \frac{\lambda^2}{2} \left( 1 + \sqrt{1 - \frac{12}{\lambda^2}} \right). \]
For $\lambda^2$ growing from $12$ to infinity, $\xi_\mathrm{max}$ decreases from $6$ to $3$ (the location of the photon sphere), and $U_\lambda(\xi_\mathrm{max})$ grows from $8/9$ to infinity. At the same time, $\xi_\mathrm{min}$ grows from $6$ to infinity, while $U_\lambda(\xi_\mathrm{min})$ grows from $8/9$ to $1$.

Of special interest is the limiting value of the angular momentum $\lambda_c(\varepsilon)$ such that $U_{\lambda_c(\varepsilon)}(\xi_\mathrm{max})$ is equal to an a priori prescribed value of $\varepsilon^2$, i.e., a solution to the equation
\[ U_{\lambda_c(\varepsilon)}(\xi_\mathrm{max}) = \varepsilon^2. \]
It can be easily computed as
\begin{eqnarray}
\lambda_c(\varepsilon) & = & \sqrt{ \frac{12}{1 - 4\left(\delta - \sqrt{\delta^2 + \delta}\right)^2} } \nonumber \\
& = & \sqrt{\frac{12}{1 - 4 \delta - 8 \delta^2 + 8 \delta \sqrt{\delta^2 + \delta}}},
\label{lambdacorig}
\end{eqnarray}
where
\[ \delta = \frac{9}{8} \varepsilon^2 - 1, \]
and this is essentially the formula given in \cite{Olivier}. Unfortunately, it turns out to be numerically unstable for large values of $\varepsilon$. In numerical applications we use another form, namely,
\begin{equation}
\label{limit1}
 \lambda_c(\varepsilon)^2 = \frac{12}{1-\frac{4}{\left(\frac{3 \varepsilon }{\sqrt{9 \varepsilon ^2-8}}+1\right)^2}}, 
 \end{equation}
which reduces the numerical instability.

Particles with $\varepsilon \ge 1$ and $\lambda \le \lambda_c(\varepsilon)$ can travel from infinity and get absorbed by the black hole. We denote the quantities referring to such particles (trajectories) with the subscript or superscript $\mathrm{(abs)}$. On the other hand, particles with sufficiently high angular momentum $\lambda$ are scattered by the centrifugal barrier, and travel back to infinity. Quantities referring to those trajectories are denoted with $\mathrm{(scat)}$. The precise characterization of the phase-space region occupied by these trajectories is slightly more complex. A minimal energy of a scattered particle depends on $\xi$, and it is given by
\begin{equation}
\label{limit2}
    \varepsilon_\mathrm{min}(\xi) = \begin{cases} \infty, & \xi \leq 3 ,\\
    \sqrt{\left(1 - \frac{2}{\xi} \right) \left(1 + \frac{1}{\xi - 3} \right)}, & 3 < \xi < 4, \\
    1, & \xi \ge 4.  \end{cases}
\end{equation}
Note that $\xi = 3$ corresponds to the location of the photon sphere. Consequently, no particle can be scattered to infinity from a location with $\xi \le 3$. The upper limit on the total angular momentum of a scattered particle reads
\begin{equation} 
\label{limit3}
\lambda_\mathrm{max}(\xi,\varepsilon) = \xi \sqrt{\frac{\varepsilon^2}{1 - \frac{2}{\xi}} - 1}. 
\end{equation}
This is actually a very simple bound following directly from the requirement that $\varepsilon^2 \ge U_\lambda(\xi)$.
The phase-space region occupied by the particles traveling from infinity and scattered off the centrifugal barrier is given by the conditions $\varepsilon_\mathrm{min}(\xi) < \varepsilon < \infty$, $\lambda_c(\varepsilon) < \lambda < \lambda_\mathrm{max}(\xi,\varepsilon)$.

\subsection{Particle current density and the accretion rate}
\label{sec:jcalculation}

In order to evaluate integrals over momenta in Eqs.\ (\ref{tmunu}) or (\ref{jgeneral}), we define a new coordinate $\chi$ so that
\[ \pi_\theta = \lambda \cos \chi, \quad \lambda_z = \lambda \sin \theta \sin \chi,  \]
and change the variables $(\pi_\theta, \lambda_z)$ to $(\lambda, \chi)$. Note that Eq.\ (\ref{pthetaformula}) is then satisfied identically. In total, we change the momentum variables from $(p_t,p_r,p_\theta,p_\varphi)$ to $(\varepsilon, m, \lambda, \chi)$, according to
\[ p_t = -m \varepsilon, \quad p_\theta = M m \lambda \cos \chi, \quad p_\varphi = M m \lambda \sin \theta \sin \chi. \]
The radial momentum $p_r$ is given as a solution to the equation
\[ g^{tt} m^2 \varepsilon^2 - 2 g^{tr} m \varepsilon p_r + g^{rr} (p_r)^2 + \frac{m^2 \lambda^2}{\xi^2} + m^2 = 0. \]

For the integration element in the momentum space we get
\[ \mathrm{dvol}_x(p) = \frac{1}{\xi^2} \frac{m^3 \lambda}{ \sqrt{\varepsilon^2 - U_\lambda (\xi)}} d \varepsilon dm d\lambda d \chi. \]

Note that
\begin{eqnarray*}
\sqrt{P_1^2 - P_0^2} & = & m \sqrt{\varepsilon^2 - 1},\\
\frac{\sqrt{P_3^2 - P_2^2}}{P_3} & = & \sqrt{1 - \sin^2 \theta \sin^2 \chi}.
\end{eqnarray*}
Also note that
\[ \epsilon_\theta \arctan \left( \frac{\lambda \cot \theta}{\sqrt{\lambda^2 - \frac{\lambda_z^2}{\sin^2 \theta}}} \right) = \arctan \left( \frac{\cot \theta}{\cos \chi}\right). \]
Thus
\[ Q^3 = \epsilon_r X(\xi,\varepsilon,\lambda) - \epsilon_r \frac{\pi}{2} - \arctan \left( \frac{\cot \theta}{\cos \chi}\right). \]
This allows one to express $F$ as
\begin{widetext}
\begin{eqnarray*}
F & = & \alpha \exp \left\{ \beta \gamma \left[ - \varepsilon - v \epsilon_r \epsilon_\theta \sqrt{\varepsilon^2 - 1} \sqrt{1 - \sin^2 \theta \sin^2 \chi} \sin \left( - \epsilon_r X(\xi,\varepsilon,\lambda) + \arctan \left( \frac{\cot \theta}{\cos \chi}\right)  \right) \right]\right\} \\
& = & \alpha \exp \left\{ \beta \gamma \left[ - \varepsilon - v  \sqrt{\varepsilon^2 - 1} \left( \epsilon_r \cos X (\xi,\varepsilon,\lambda) \cos \theta - \sin X (\xi, \varepsilon,\lambda) \sin \theta \cos \chi \right) \right]\right\},
\end{eqnarray*}
where we have used the fact that $\epsilon_\theta = \mathrm{sign}(\cos \chi)$. Note that the above expression shows the correct asymptotic behavior for $X \to 0$ (or $\xi \to \infty$). 

In the following we define also
\begin{eqnarray*}
\mathcal F_n & = & \int_0^\infty m^n f dm  = m^n F \\
& = & \alpha m^n \exp \left\{ \beta \gamma \left[ - \varepsilon - v  \sqrt{\varepsilon^2 - 1} \left( \epsilon_r \cos X (\xi,\varepsilon,\lambda) \cos \theta - \sin X (\xi, \varepsilon,\lambda) \sin \theta \cos \chi \right) \right]\right\}.
\end{eqnarray*}
We abuse the notation at this point: the symbol $m$ is used to denote the mass understood as one of the phase-space coordinates, and also as the fixed mass of the gas particles. While using two different symbols in this context would be mathematically more clear, it could turn out to be physically confusing.

We are now ready to compute the components of the particle current density $J_\mu$, which we express as a sum of two parts, $J_\mu = J_\mu^\mathrm{(abs)} + J_\mu^\mathrm{(scat)}$, corresponding to absorbed and scattered particles, respectively. The components $J_\mu^\mathrm{(abs)}$ are computed as
\[ J_\mu^\mathrm{(abs)} = \int_1^\infty d\varepsilon \int_0^{\lambda_c(\varepsilon)} d \lambda \int_0^{2 \pi} d \chi \int_0^\infty dm \frac{ p_\mu f m^3 \lambda}{ \xi^2 \sqrt{\varepsilon^2 - U_\lambda (\xi)}}, \]
where we have to keep in mind that the absorbed trajectories correspond to $\epsilon_r = -1$ ($\epsilon_r$ appears in the expressions for $p_r$ and $f$). For the components $J_\mu^\mathrm{(scat)}$ we have
\[  J_\mu^\mathrm{(scat)} = \sum_{\epsilon_r = \pm 1} \int_{\varepsilon_\mathrm{min}(\xi)}^\infty d\varepsilon \int_{\lambda_c(\varepsilon)}^{\lambda_\mathrm{max}(\xi,\varepsilon)} d \lambda \int_0^{2 \pi} d \chi \int_0^\infty dm \frac{ p_\mu f m^3 \lambda}{ \xi^2 \sqrt{\varepsilon^2 - U_\lambda (\xi)}}. \]
The components of the four-momentum can be expressed as
\[ p_\mu = (p_t, p_r, p_\theta, p_\varphi) = m \left( - \varepsilon, \frac{ (1 - N \eta) \varepsilon + \epsilon_r \sqrt{\varepsilon^2 - U_\lambda (\xi)}}{N} , M \lambda \cos \chi, M \lambda \sin \theta \sin \chi \right). \]
To get $J_t^\mathrm{(abs)}$ we evaluate
\begin{eqnarray}
J_t^\mathrm{(abs)} & = & - \frac{1}{\xi^2} \int_1^\infty d \varepsilon \int_0^{\lambda_c(\varepsilon)} d \lambda \int_0^{2 \pi} d \chi \varepsilon \mathcal F_4 \frac{\lambda}{ \sqrt{\varepsilon^2 - U_\lambda (\xi)}} \nonumber \\
& = & -\frac{2 \pi \alpha m^4}{\xi^2} \int_1^\infty d \varepsilon \varepsilon e^{- \beta \gamma \varepsilon} \int_0^{\lambda_c(\varepsilon)} d \lambda \frac{\lambda}{\sqrt{\varepsilon^2 - U_\lambda (\xi)}} \exp \left[ \beta \gamma v \sqrt{\varepsilon^2 - 1} \cos X \cos \theta \right] \nonumber \\
& & \times I_0\left[\beta \gamma v \sqrt{\varepsilon^2 - 1} \sin X \sin \theta \right],
\label{jtabs}
\end{eqnarray}
where we have used the fact that
\[ \int_0^{2 \pi} e^{\delta \cos \chi} d \chi = 2 \pi I_0(\delta), \]
and $I_n(\delta)$ is the modified Bessel function of the first kind.
For scattered particles we get
\begin{eqnarray}
J_t^\mathrm{(scat)} & = & - \frac{1}{\xi^2} \sum_{\epsilon_r = \pm 1} \int_{\varepsilon_\mathrm{min}(\xi)}^\infty d \varepsilon \int_{\lambda_c(\varepsilon)}^{\lambda_\mathrm{max}(\xi,\varepsilon)} d \lambda \int_0^{2 \pi} d \chi \varepsilon \mathcal F_4 \frac{\lambda}{ \sqrt{\varepsilon^2 - U_\lambda (\xi)}} \nonumber \\
& = & - \frac{4 \pi \alpha m^4}{\xi^2} \int_{\varepsilon_\mathrm{min}(\xi)}^\infty d \varepsilon \varepsilon e^{- \beta \gamma \varepsilon} \int_{\lambda_c(\varepsilon)}^{\lambda_\mathrm{max}(\xi,\varepsilon)} d \lambda \frac{\lambda}{\sqrt{\varepsilon^2 - U_\lambda(\xi)}} \cosh \left( \beta \gamma v \sqrt{\varepsilon^2 - 1} \cos X \cos \theta \right) \nonumber  \\
& & \times I_0 \left( \beta \gamma v \sqrt{\varepsilon^2 - 1} \sin X \sin \theta \right).
\label{Jtscat}
\end{eqnarray}

In evaluating $J_r^\mathrm{(abs)}$, one should note that an expression for $p_r$ with $\epsilon_r = -1$ which is manifestly regular at the horizon can be written as
\[ p_r = m \pi_\xi = m \left( - \eta \varepsilon + \frac{1 + \frac{\lambda^2}{\xi^2}}{\varepsilon + \sqrt{\varepsilon^2 - U_\lambda(\xi)}} \right). \]
For $\epsilon_r = +1$ the component $p_r$ is divergent at the horizon. Note that this does not cause any difficulties, as $J_\mu^\mathrm{(scat)}$ vanishes below the photon sphere (for $\xi < 3$), and $p_r$ with $\epsilon_r = +1$ only appears in expressions for $J_\mu^\mathrm{(scat)}$. The formulas for $J_r$ and $J_\theta$ are as follows:
\begin{subequations}
\label{jrjtheta}
\begin{eqnarray}
J_r^\mathrm{(abs)} & = & \frac{2 \pi \alpha m^4}{\xi^2} \int_1^\infty d \varepsilon \; e^{- \beta \gamma \varepsilon} \int_0^{\lambda_c(\varepsilon)} d \lambda \frac{\lambda}{\sqrt{\varepsilon^2 - U_\lambda (\xi)}} \left( - \eta \varepsilon + \frac{1 + \frac{\lambda^2}{\xi^2}}{\varepsilon + \sqrt{\varepsilon^2 - U_\lambda (\xi)}} \right) \nonumber \\
& & \times \exp \left[ \beta \gamma v \sqrt{\varepsilon^2 - 1} \cos X \cos \theta \right] I_0\left[\beta \gamma v \sqrt{\varepsilon^2 - 1} \sin X \sin \theta \right], \\
J_r^\mathrm{(scat)} & = & \frac{4 \pi \alpha m^4}{\xi^2 N} \int_{\varepsilon_\mathrm{min}(\xi)}^\infty d \varepsilon e^{- \beta \gamma \varepsilon} \int_{\lambda_c(\varepsilon)}^{\lambda_\mathrm{max}(\xi,\varepsilon)} d \lambda \lambda \left[ \frac{(1 - N \eta) \varepsilon}{\sqrt{\varepsilon^2 - U_\lambda}} \cosh \left(\beta \gamma v \sqrt{\varepsilon^2 - 1} \cos X \cos \theta \right) \right. \nonumber \\
& & \left. - \sinh \left(\beta \gamma v \sqrt{\varepsilon^2 - 1} \cos X \cos \theta \right) \right]  I_0\left[\beta \gamma v \sqrt{\varepsilon^2 - 1} \sin X \sin \theta \right] \\
J_\theta^\mathrm{(abs)} & = & \frac{2 \pi \alpha M m^4}{\xi^2} \int_1^\infty d \varepsilon e^{- \beta \gamma \varepsilon} \int_0^{\lambda_c(\varepsilon)} d \lambda \frac{\lambda^2}{\sqrt{\varepsilon^2 - U_\lambda (\xi)}} \exp \left( \beta \gamma v \sqrt{\varepsilon^2 - 1} \cos X \cos \theta \right) \nonumber \\
&& \times I_1 \left(\beta \gamma v \sqrt{\varepsilon^2 - 1} \sin X \sin \theta \right), \\
J_\theta^\mathrm{(scat)} & = & - \frac{4 \pi \alpha M m^4}{\xi^2} \int_{\varepsilon_\mathrm{min}(\xi)}^\infty d \varepsilon e^{- \beta \gamma \varepsilon} \int_{\lambda_c(\varepsilon)}^{\lambda_\mathrm{max}(\xi,\varepsilon)} d \lambda \frac{\lambda^2}{\sqrt{\varepsilon^2 - U_\lambda (\xi)}} \cosh \left( \beta \gamma v \sqrt{\varepsilon^2 - 1} \cos X \cos \theta \right) \nonumber \\
&& \times I_1 \left( \beta \gamma v \sqrt{\varepsilon^2 - 1} \sin X \sin \theta \right).
\end{eqnarray}
\end{subequations}
In computing $J_\theta$, we make use of the integral
\[ \int_0^{2 \pi} \cos \chi e^{\delta \cos \chi} d \chi = 2 \pi I_1 (\delta). \]
Finally, since
\[ \int_0^{2 \pi} \sin \chi e^{\delta \cos \chi} d \chi = 0, \]
we obtain $J_\varphi = 0$. 

We will now discuss the mass accretion rate through a sphere of a given radius $r = M \xi$, defined as
\begin{equation} 
\label{Mdot_general}
\dot M = - m \int_0^{2 \pi} d\varphi \int_0^\pi d \theta\; r^2 \sin \theta \; J^r. 
\end{equation}
It follows immediately from Eq.\ (\ref{conservationj}) that $\dot M$ does not depend on the radius of the sphere. Taking into account that $J^r = g^{rr}J_r + g^{rt}J_t$, or $p^r = g^{rr}p_r + g^{rt}p_t = \epsilon_r m \sqrt{\varepsilon^2 - U_\lambda (\xi)}$, we get in general
\[ J^r = \frac{1}{\xi^2} \int d \varepsilon \int d \lambda \int d \chi \epsilon_r \lambda \mathcal F_4. \]
A straightforward calculation yields
\[ J^r_\mathrm{(abs)} = - \frac{2 \pi \alpha m^4}{\xi^2} \int_1^\infty d \varepsilon \; e^{- \beta \gamma \varepsilon} \int_0^{\lambda_c(\varepsilon)} d \lambda \; \lambda \exp \left(\beta \gamma v \sqrt{\varepsilon^2 - 1} \cos X \cos \theta \right) I_0 \left( \beta \gamma v \sqrt{\varepsilon^2 - 1} \sin X \sin \theta \right) \]
and
\[ J^r_\mathrm{(scat)} = - \frac{4 \pi \alpha m^4}{\xi^2} \int_{\varepsilon_\mathrm{min}(\xi)}^\infty d \varepsilon \; e^{- \beta \gamma \varepsilon} \int_{\lambda_c(\varepsilon)}^{\lambda_\mathrm{max}(\xi,\varepsilon)} d \lambda \; \lambda \sinh\left( \beta \gamma v \sqrt{\varepsilon^2 - 1} \cos X \cos \theta \right) I_0 \left( \beta \gamma v \sqrt{\varepsilon^2 - 1} \sin X \sin \theta \right). \]
It is easy to see that 
$J^r_\mathrm{(scat)}$ does not contribute to the mass accretion rate. Indeed,
\[ \int_0^\pi d \theta \sin \theta J^r_\mathrm{(scat)} = 0. \]
This is because
\[ \int_0^\pi I_0 (\beta \sin \theta) \sinh (\delta \cos \theta) \sin \theta d \theta = \int_{-1}^1 I_0 \left(\beta \sqrt{1 - x^2} \right) \sinh(\delta x)  dx = 0, \]
where $x = \cos \theta$. This is also reasonable, as we could try to evaluate $\dot M$ at a sphere with a radius $\xi < 3$, where $J^r_\mathrm{(scat)} \equiv 0$. In order to facilitate the computation of $\dot M$ we first take the limit of $\xi \to \infty$ ($X \to 0$). In this limit
\[ \xi^2 J^r_\mathrm{(abs)} \to - \pi \alpha m^4 \int_1^\infty d \varepsilon \lambda_c^2(\varepsilon) e^{- \beta \gamma \varepsilon}  e^{\beta \gamma v \sqrt{\varepsilon^2 - 1} \cos \theta}. \]
Consequently, $\dot M$ (evaluated at infinity) reads
\begin{eqnarray}
\dot M & = & 4 \pi^2 \alpha M^2 m^5 \int_1^\infty d \varepsilon \; e^{-\beta \gamma \varepsilon} \lambda_c^2(\varepsilon) \frac{\sinh \left( \beta \gamma v \sqrt{\varepsilon^2 - 1}  \right)}{\beta \gamma v \sqrt{\varepsilon^2 - 1}} \nonumber \\
& = & \pi M^2 m n_\infty \frac{\beta}{K_2(\beta)}  \int_1^\infty d \varepsilon \; e^{-\beta \gamma \varepsilon} \lambda_c^2(\varepsilon) \frac{\sinh \left( \beta \gamma v \sqrt{\varepsilon^2 - 1}  \right)}{\beta \gamma v \sqrt{\varepsilon^2 - 1}}.
\label{Mdot_infty}
\end{eqnarray}
\end{widetext}
Note that for $v = 0$ we recover the result of \cite{cieslik_mach} [Eq.\ (40) of \cite{cieslik_mach}]. As yet another consistency test, we checked that expression (\ref{Mdot_general}) evaluated numerically at spheres with different radii $\xi$ agrees with $\dot M$ given by Eq.\ (\ref{Mdot_infty}) with an accuracy of at least $10^{-4}$.

In the limit of $\beta \to 0$ (infinite temperature, ultrarelativistic gas) we found
\begin{equation}
\label{limitbeta0}
\dot M = 27 \pi M^2 m  n_\infty  \gamma.
\end{equation}
This agrees qualitatively with the result of Petrich, Shapiro, and Teukolsky \cite{petrich}, who found for the perfect gas with the ultrahard equation of state the expression $\dot M = 16 \pi M^2 m  n_\infty  \gamma$. Similarly to the result derived in \cite{petrich}, for large temperatures the accretion rate is essentially the value for $v = 0$ multiplied by the Lorentz factor. For $v = 0$, it is also possible to compute the low-temperature limit ($\beta \to \infty$). We get in this case
\[ \dot{M} (v=0)  =  16 M^2 m n_\infty \sqrt{2 \pi \beta} \]
(this result has already been derived in \cite{Olivier}).
For $v > 0$ we get in the low-temperature limit an approximate expression
\begin{equation}
\label{lowtemperaturelimit}
\dot M = \pi M^2 m n_\infty \frac{\lambda_c(\gamma)^2}{\gamma v},
\end{equation}
which has a local minimum for $v = \sqrt{3/8}$ and diverges for $v = 0$ and $v \to 1$. A detailed derivation of the above limits is given in Appendix \ref{appendixb}.

Another quantity that can be used to characterize the accretion is the energy accretion rate $\dot {\mathcal E}$. It can be computed in a way similar to $\dot M$, taking into account the conservation law $\nabla_\mu (T\indices{^\mu_\nu} k^\nu) = 0$, instead of $\nabla_\mu J^\mu = 0$, i.e., replacing $J^\mu$ with $J_\varepsilon^\mu = - T\indices{^\mu_\nu} k^\nu = - T\indices{^\mu_t}$. The energy accretion rate is then defined as
\[ \dot {\mathcal E} = \int_0^{2\pi} d \varphi \int_0^\pi d \theta r^2 \sin \theta \, T\indices{^r_t}. \]
The component $T\indices{^r_t}$ can be computed according to Eq.\ (\ref{tmunu}), again by splitting into the parts corresponding to scattered and absorbed trajectories. As with $\dot M$, one can show that only the latter part contributes to the energy accretion rate.
The final expression for $\dot {\mathcal E}$ reads
\begin{widetext}
\begin{eqnarray}
\dot {\mathcal E} & = & 4 \pi^2 \alpha M^2 m^5 \int_1^\infty d \varepsilon \; e^{-\beta \gamma \varepsilon} \varepsilon \lambda_c^2(\varepsilon) \frac{\sinh \left( \beta \gamma v \sqrt{\varepsilon^2 - 1}  \right)}{\beta \gamma v \sqrt{\varepsilon^2 - 1}} \nonumber \\
& = & \pi M^2 m n_\infty \frac{\beta}{K_2(\beta)}  \int_1^\infty d \varepsilon \; e^{-\beta \gamma \varepsilon} \varepsilon \lambda_c^2(\varepsilon) \frac{\sinh \left( \beta \gamma v \sqrt{\varepsilon^2 - 1}  \right)}{\beta \gamma v \sqrt{\varepsilon^2 - 1}}.
\label{Edot_infty}
\end{eqnarray}
Again, for $v = 0$ the above expression coincides with the result derived in \cite{Olivier,Olivier2}. In physical applications it is probably more reasonable to normalize $\dot {\mathcal E}$ by the asymptotic energy density $\varepsilon_\infty$. For the Maxwell-J\"{u}ttner distribution in the Minkowski spacetime it can be defined as $\varepsilon_\infty = T_{\mu\nu}k^\mu k^\nu = T_{tt} = - T\indices{^t_t}$. A direct calculation yields \cite{Olivier,israel}
\[ \varepsilon_\infty = 4 \pi \alpha m^5 \left[ \frac{K_1(\beta)}{\beta} + 3 \frac{K_2(\beta)}{\beta^2} \right]. \]
Thus
\begin{eqnarray}
\dot {\mathcal E} & = & \pi M^2 \frac{\varepsilon_\infty}{3 + \frac{\beta K_1(\beta)}{K_2(\beta)}} \frac{\beta^2}{K_2(\beta)} \int_1^\infty d \varepsilon \; e^{-\beta \gamma \varepsilon} \varepsilon \lambda_c^2(\varepsilon) \frac{\sinh \left( \beta \gamma v \sqrt{\varepsilon^2 - 1}  \right)}{\beta \gamma v \sqrt{\varepsilon^2 - 1}}.
\label{Edot_infty_epsilon}
\end{eqnarray}
\end{widetext}
The reasoning analogous to that described in Appendix \ref{appendixb} [it is possible to provide both lower and upper estimates of the integral in Eq.\ (\ref{Edot_infty_epsilon})] shows that in the limit of $\beta \to 0$ one obtains
\[ \dot {\mathcal E} = 27 \pi M^2 \varepsilon_\infty \frac{4 \gamma^2 - 1}{3}, \]
meaning that the energy accretion rate is essentially proportional to the square of the Lorentz factor $\gamma$.

\section{Numerical results}
\label{sec:numeric}

\begin{figure}
\centering
    \includegraphics[width=\columnwidth]{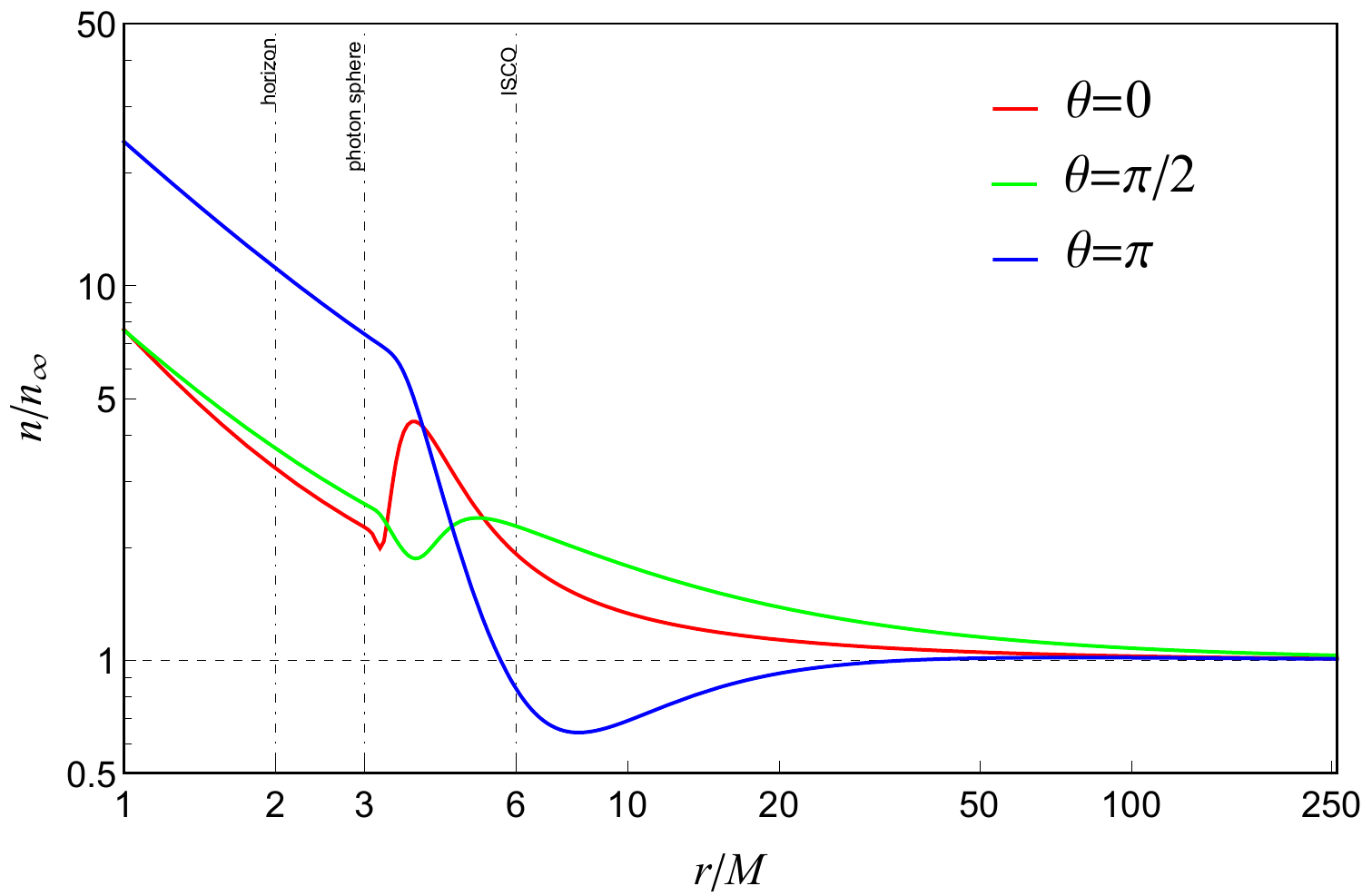}\\\includegraphics[width=\columnwidth]{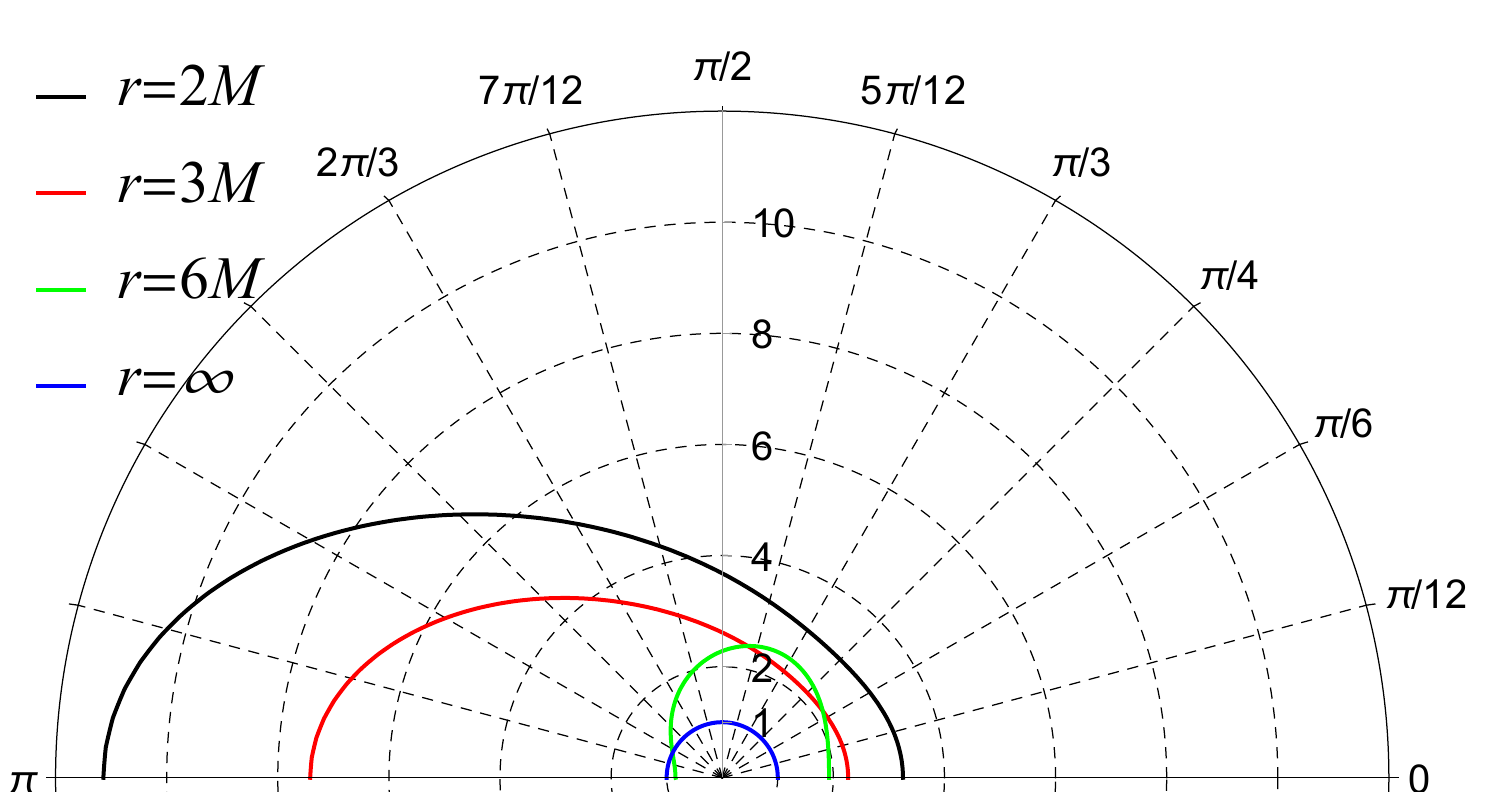}
 \caption{\label{fig:n} Typical structure of the particle density ratio $n/n_\infty$ for $\beta=8$ and $v=0.5$. The black hole moves towards $\theta = 0$. The upper panel shows the radial profiles of the density for three selected values of the angle $\theta$. The lower panel depicts angular profiles of the ratio $n/n_\infty$ computed for a sample of spheres of constant radius $\xi$. Vertical lines in the upper plot correspond to the locations of the horizon ($r/M = 2$), the photon sphere ($r/M = 3$), and the innermost stable circular orbit ($r/M = 6$).}
\end{figure}

\begin{figure*}
\centering
   \includegraphics[width=0.8\linewidth]{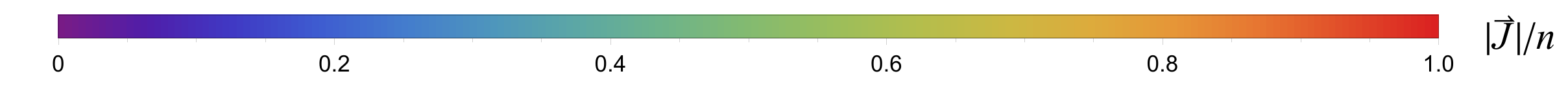}\\
    \includegraphics[width=0.666\columnwidth]{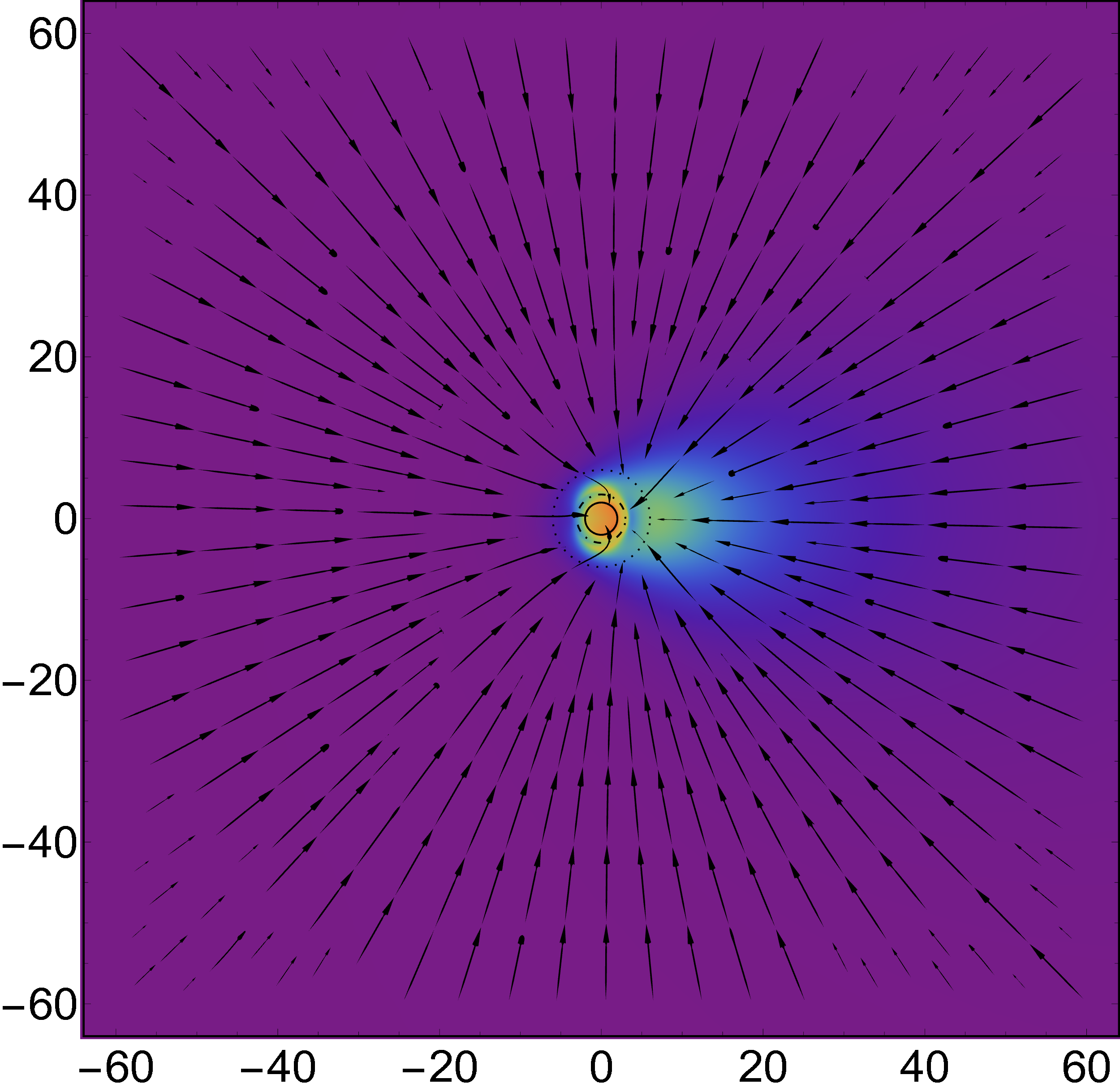}~\includegraphics[width=0.666\columnwidth]{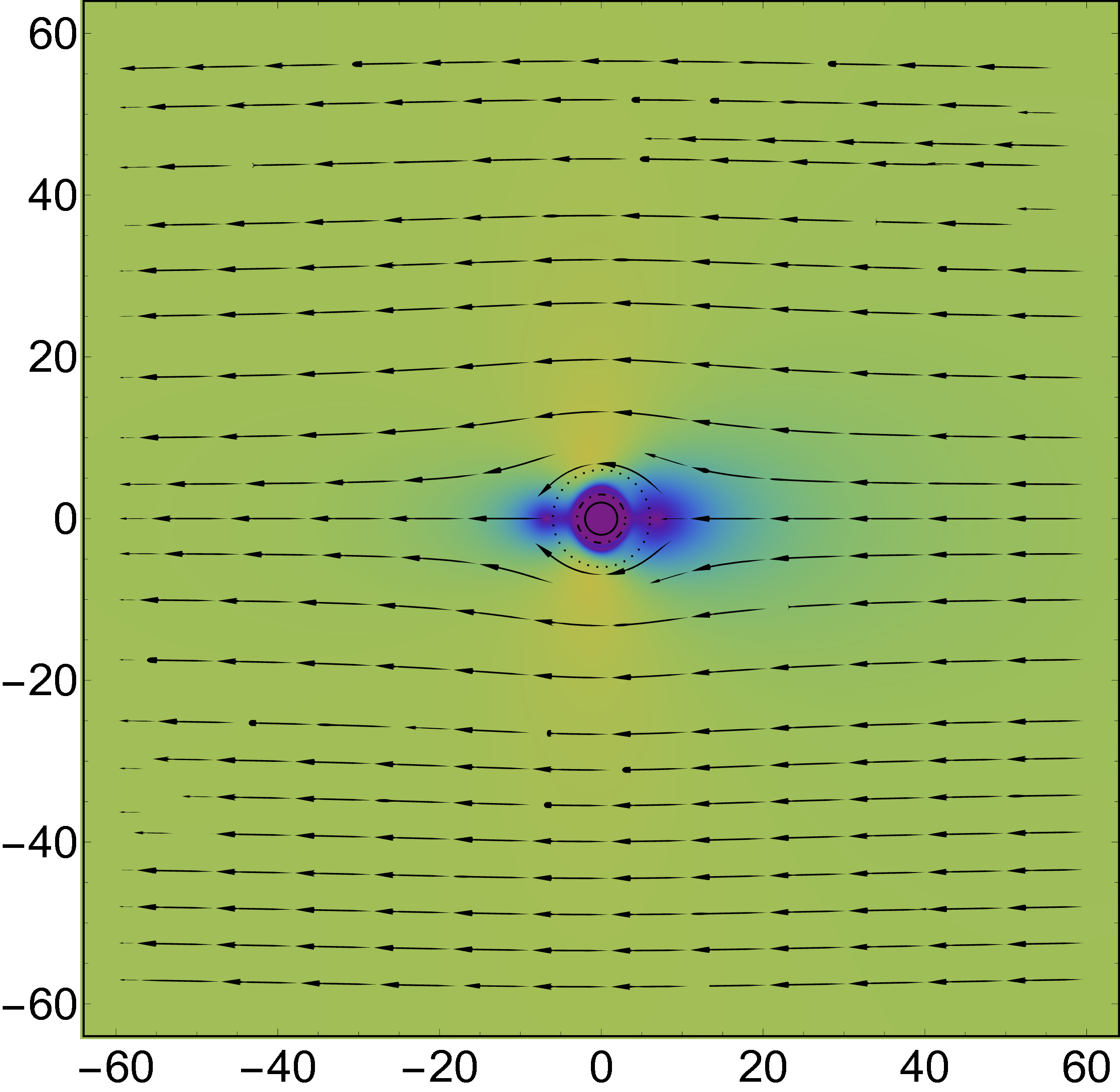}~\includegraphics[width=0.666\columnwidth]{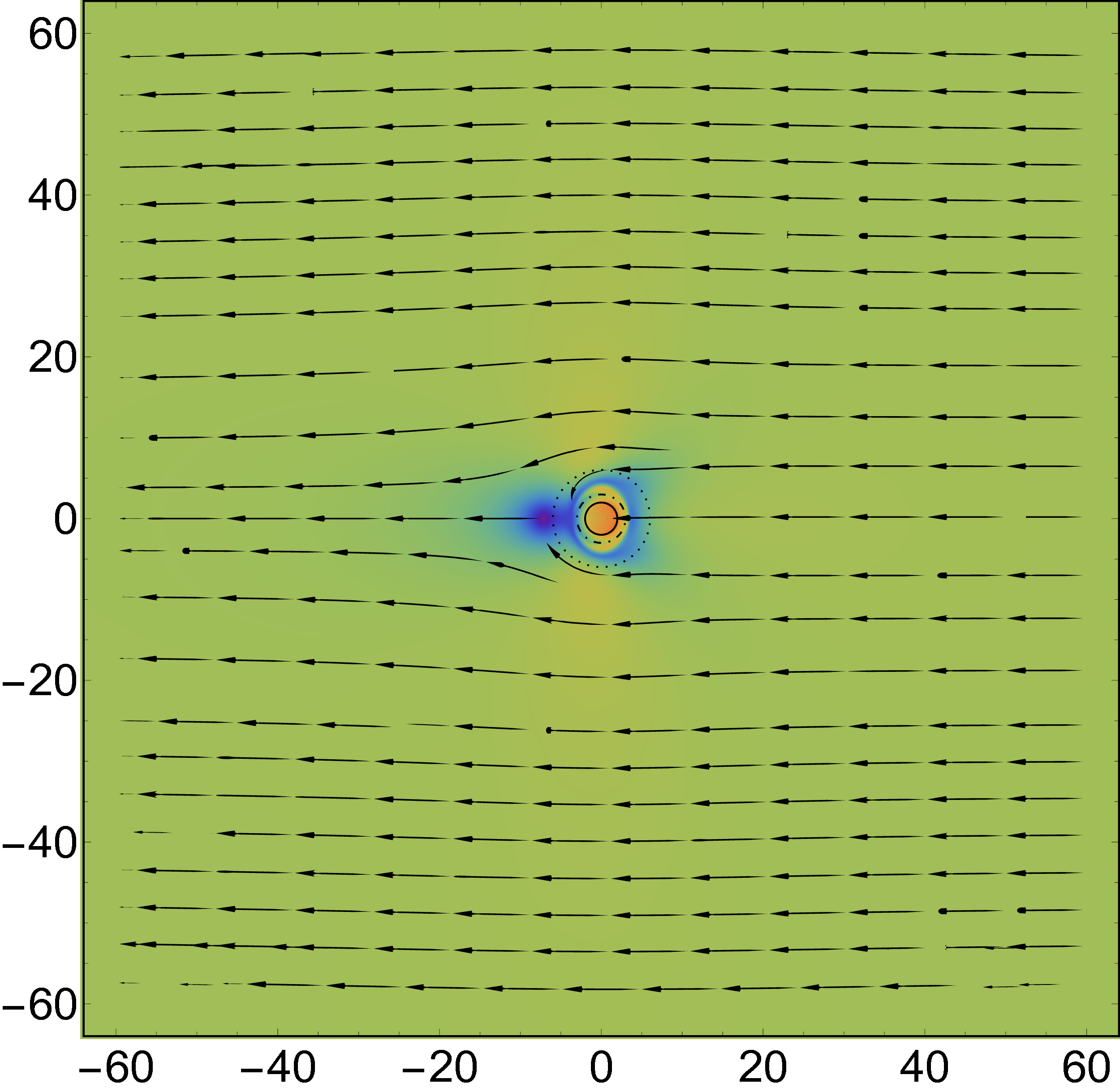}\\
    \includegraphics[width=0.666\columnwidth]{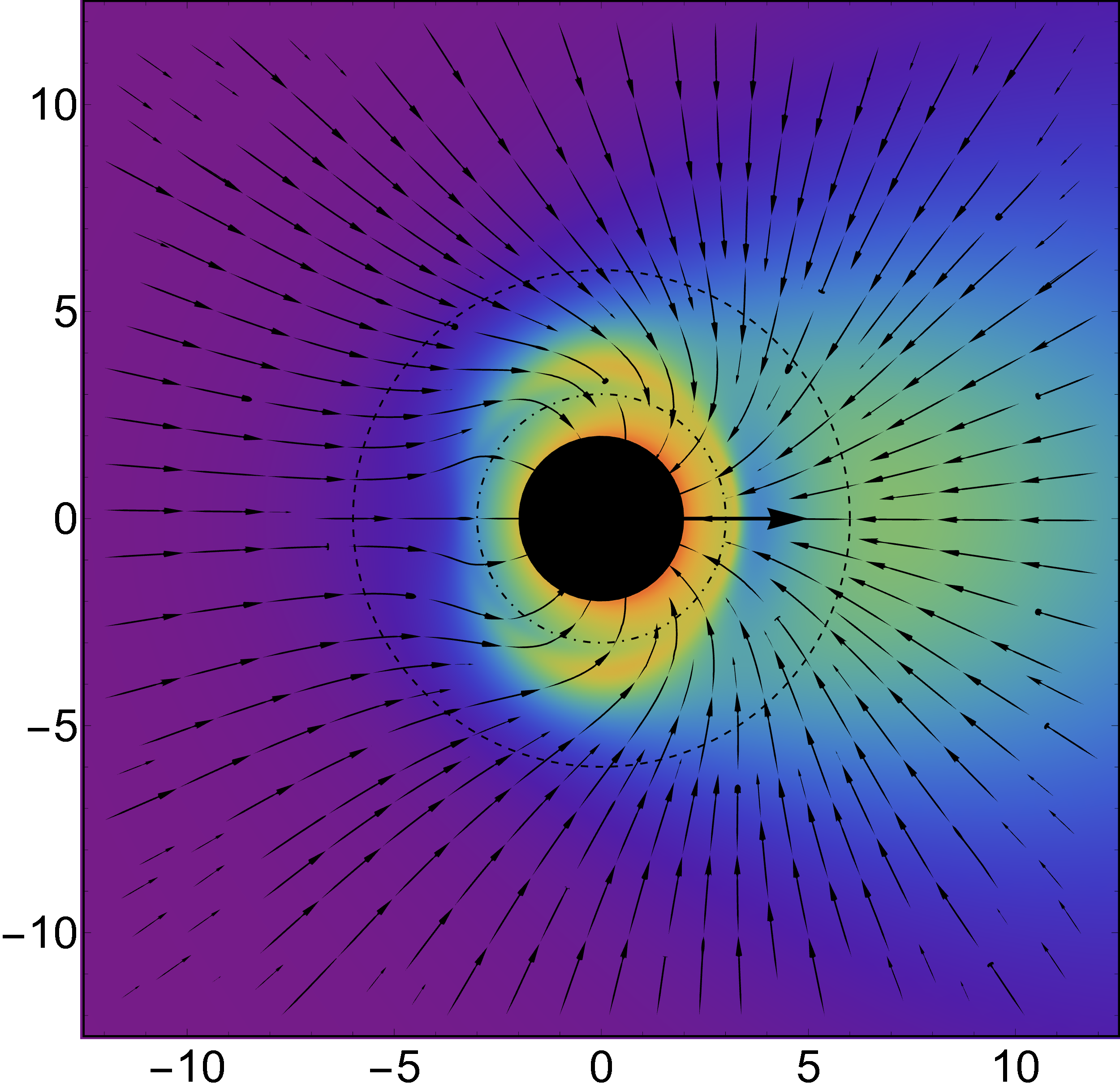}~\includegraphics[width=0.666\columnwidth]{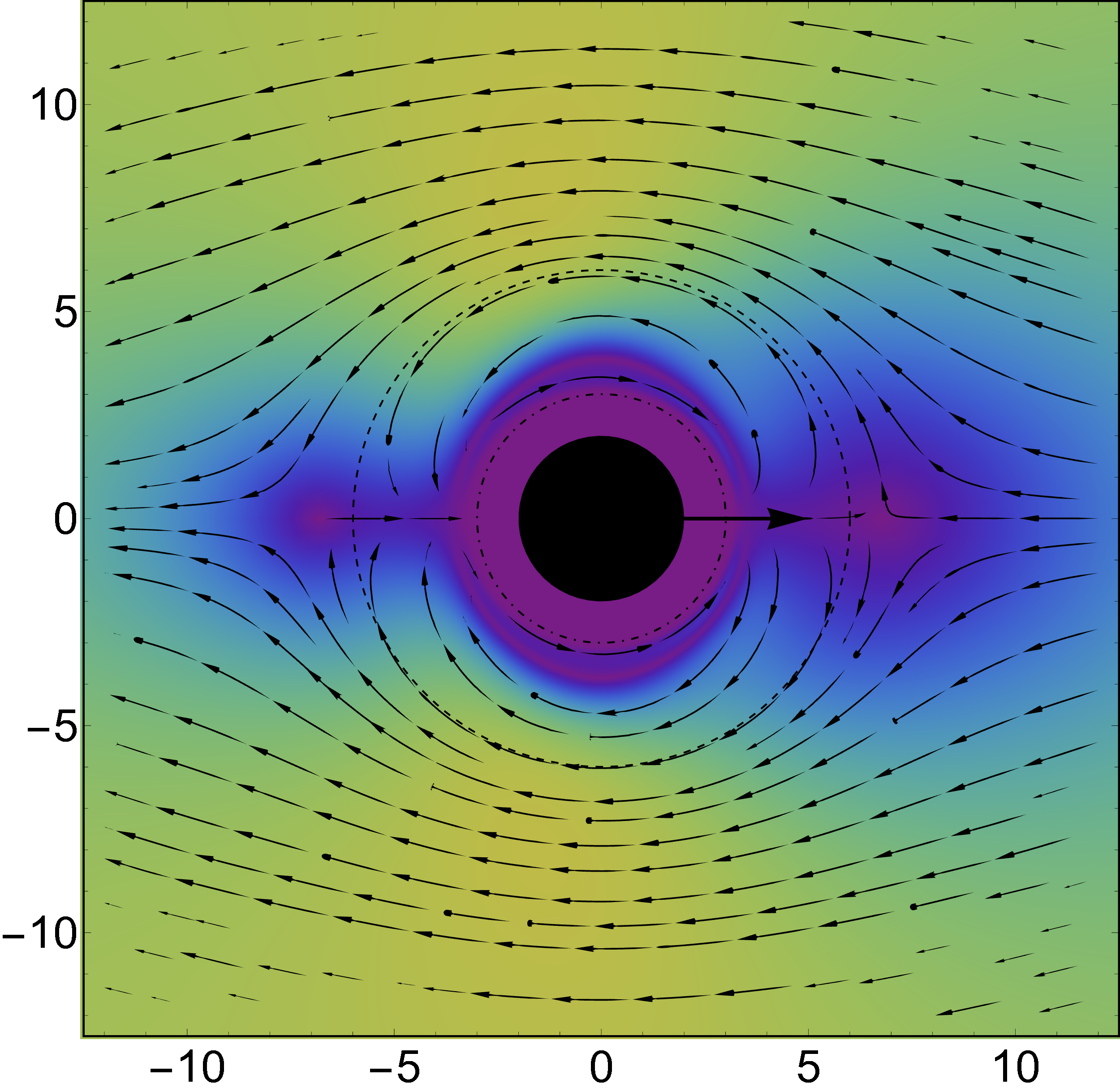}~\includegraphics[width=0.666\columnwidth]{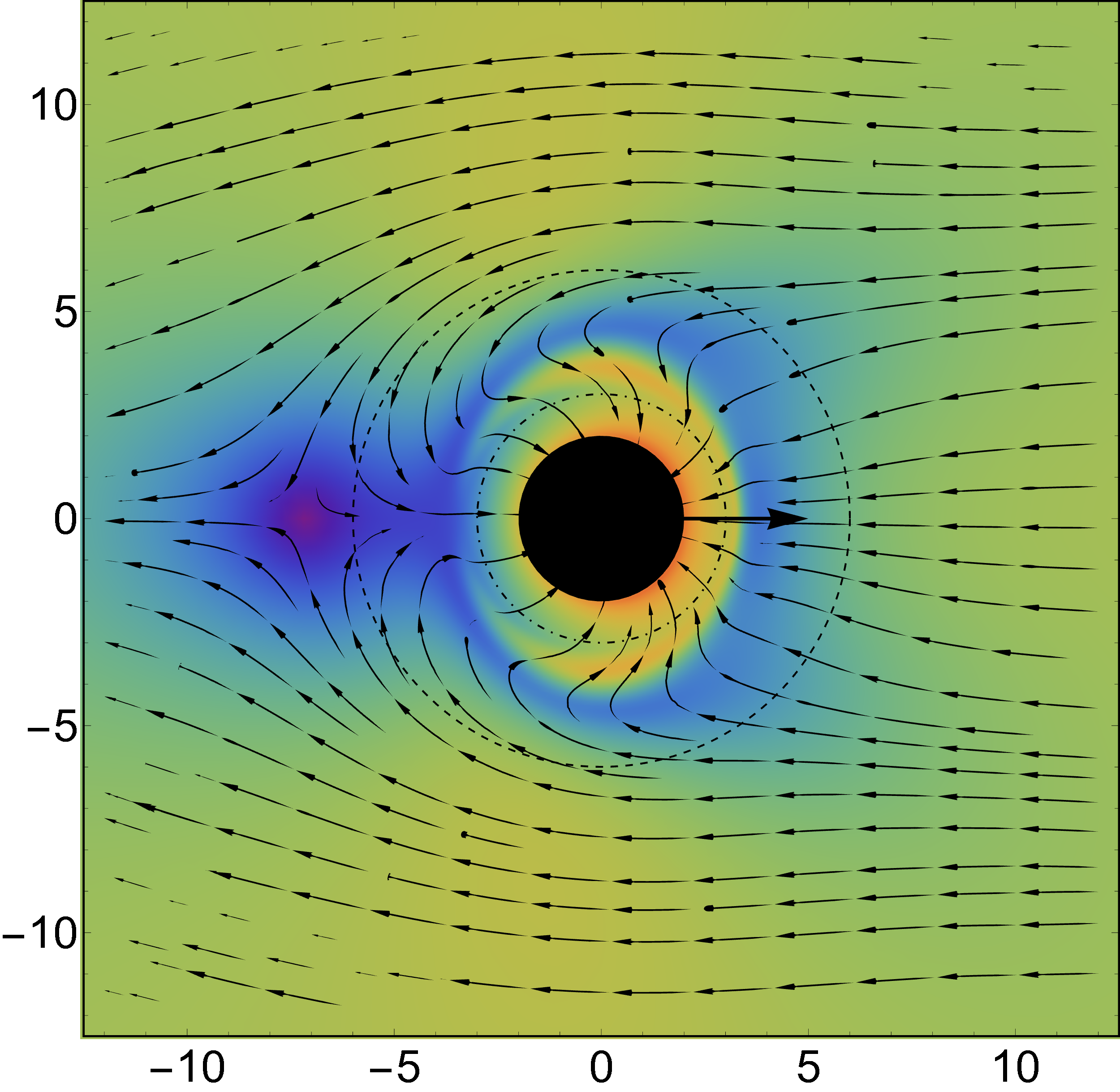}\\
    
 \caption{\label{fig:AbsvsScattvsTotal} Typical structure of the particle density current $J^\mu_\text{(abs)}$ (left column), $J^\mu_\text{(scat)}$ (middle column), and the total $J^\mu = J^\mu_\text{(abs)} + J^\mu_\text{(scat)}$ (right column) for $v = 0.5$ and $\beta = 8$. The directions of the vector field $(J^y,J^z)$ are shown with arrows and streamlines. The colors on the plot are proportional to the ratio $|\vec J|/n$.
 }
\end{figure*}

\begin{figure*}
    \centering
    \includegraphics[width=0.8\linewidth]{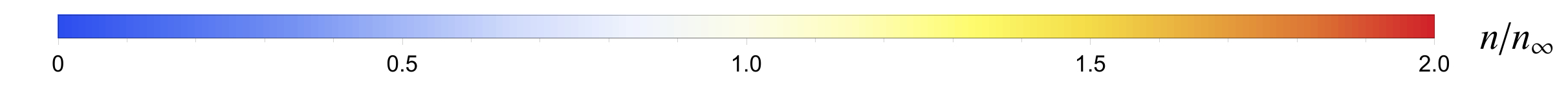}
    \includegraphics[width=0.666\columnwidth]{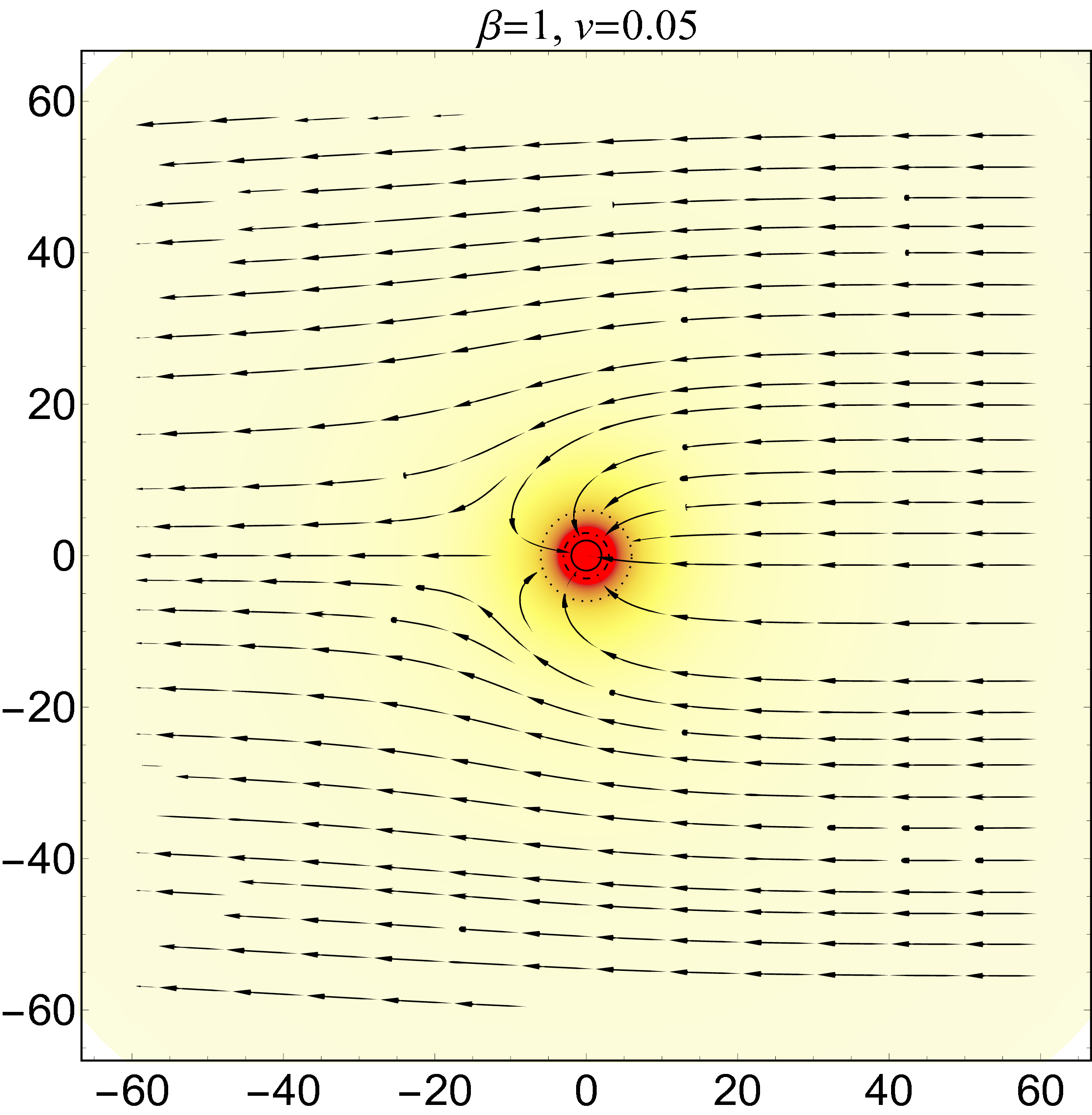}~\includegraphics[width=0.666\columnwidth]{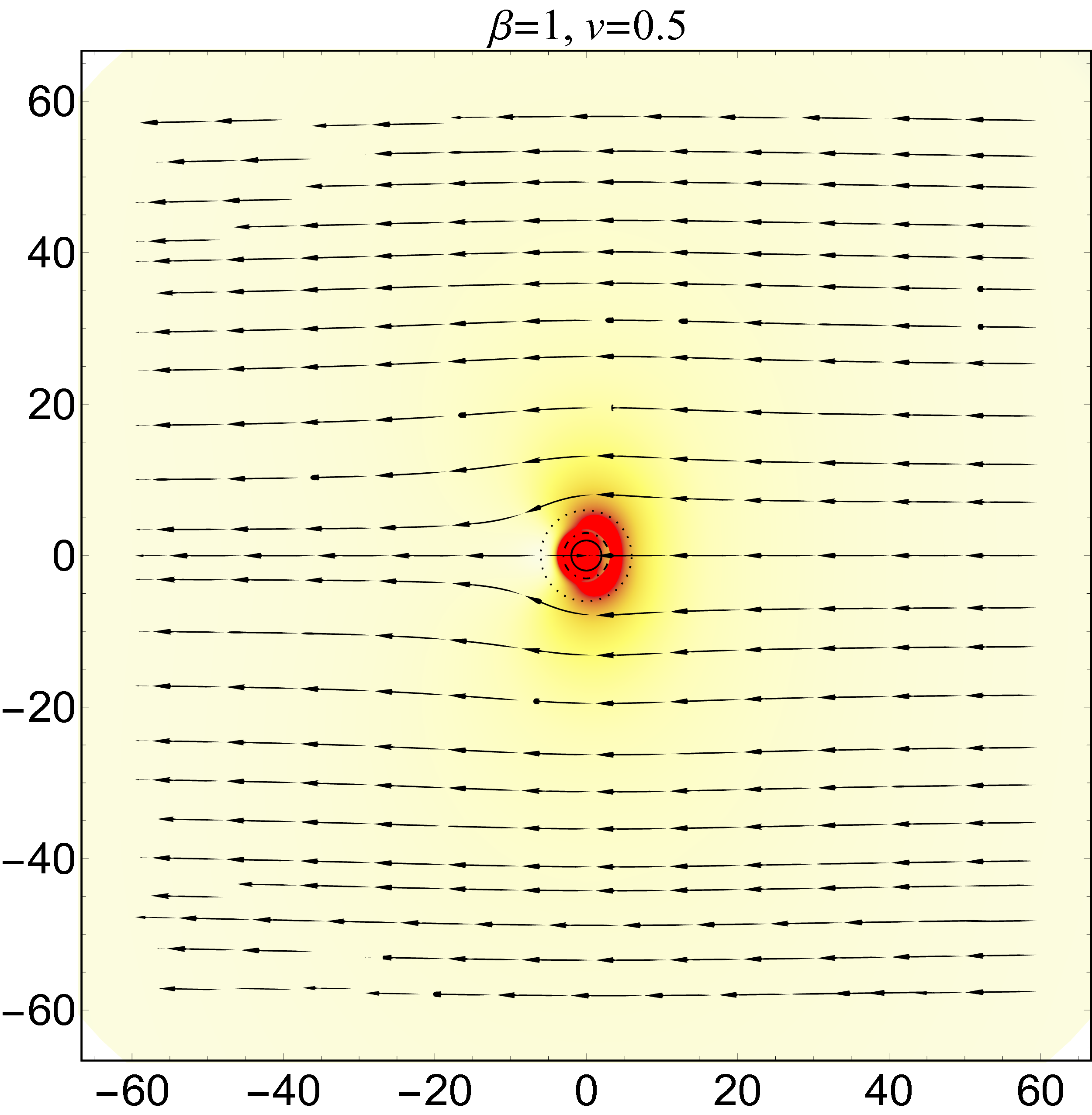}~\includegraphics[width=0.666\columnwidth]{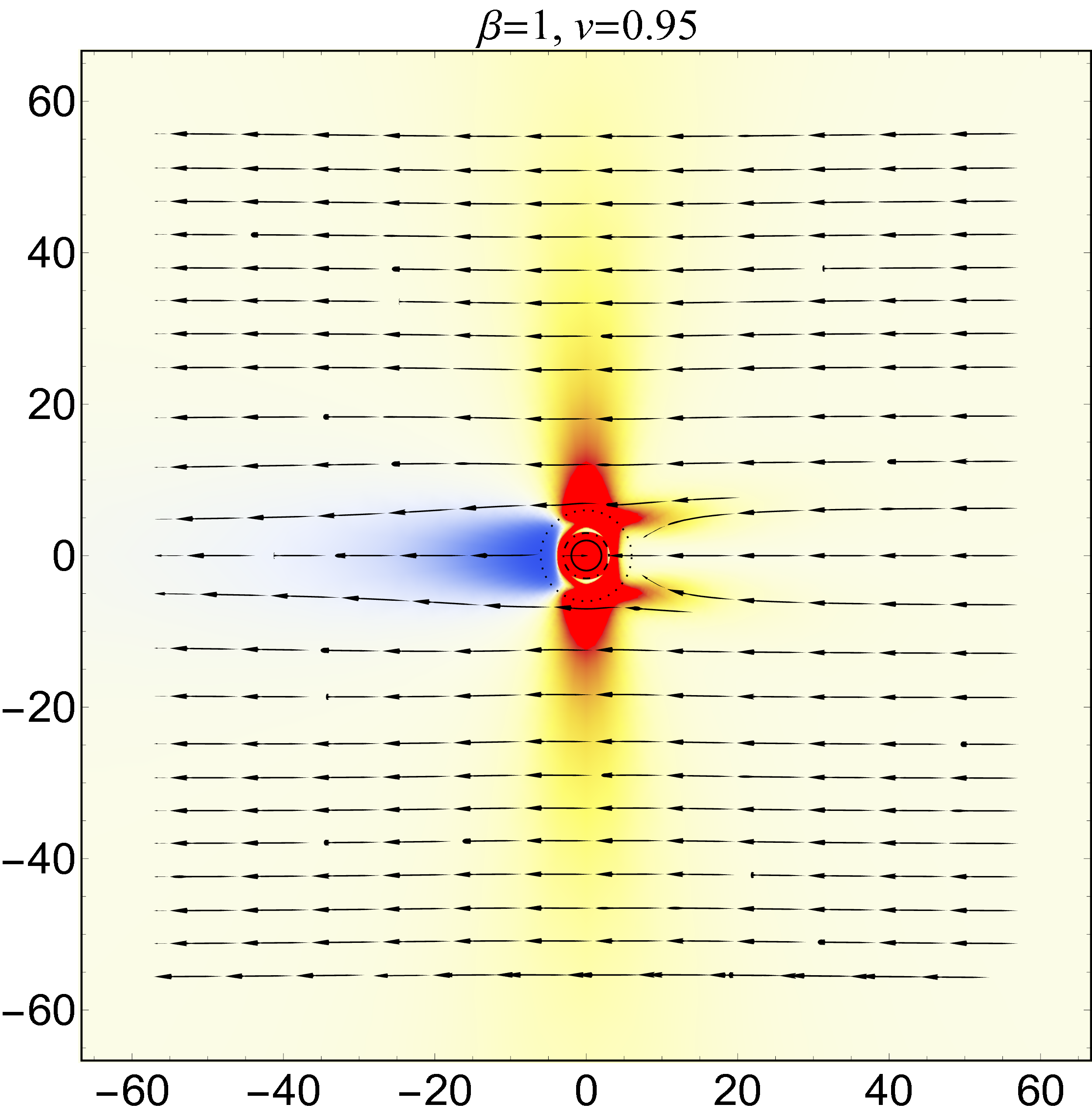}\\
    \includegraphics[width=0.666\columnwidth]{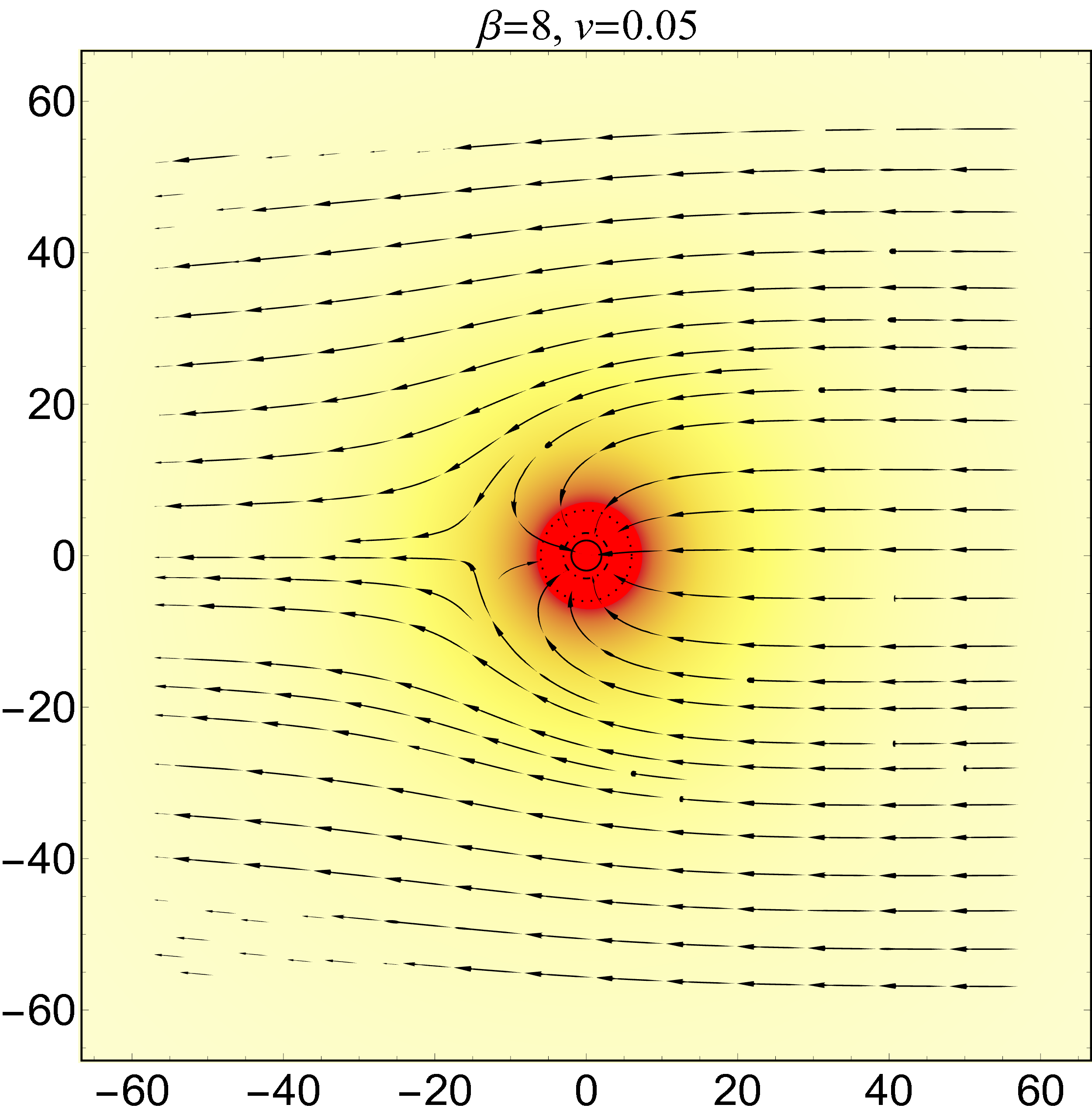}~\includegraphics[width=0.666\columnwidth]{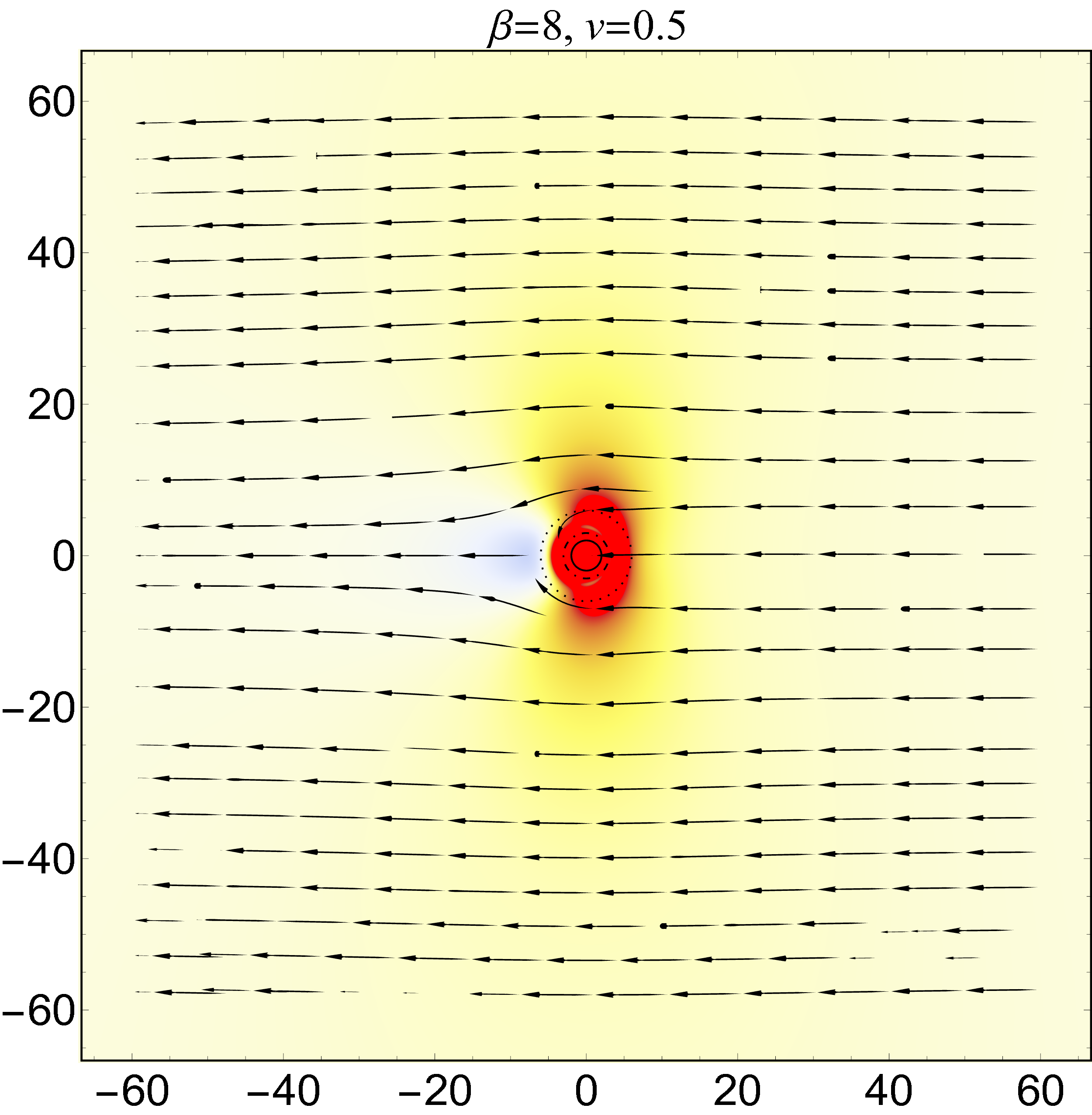}~\includegraphics[width=0.666\columnwidth]{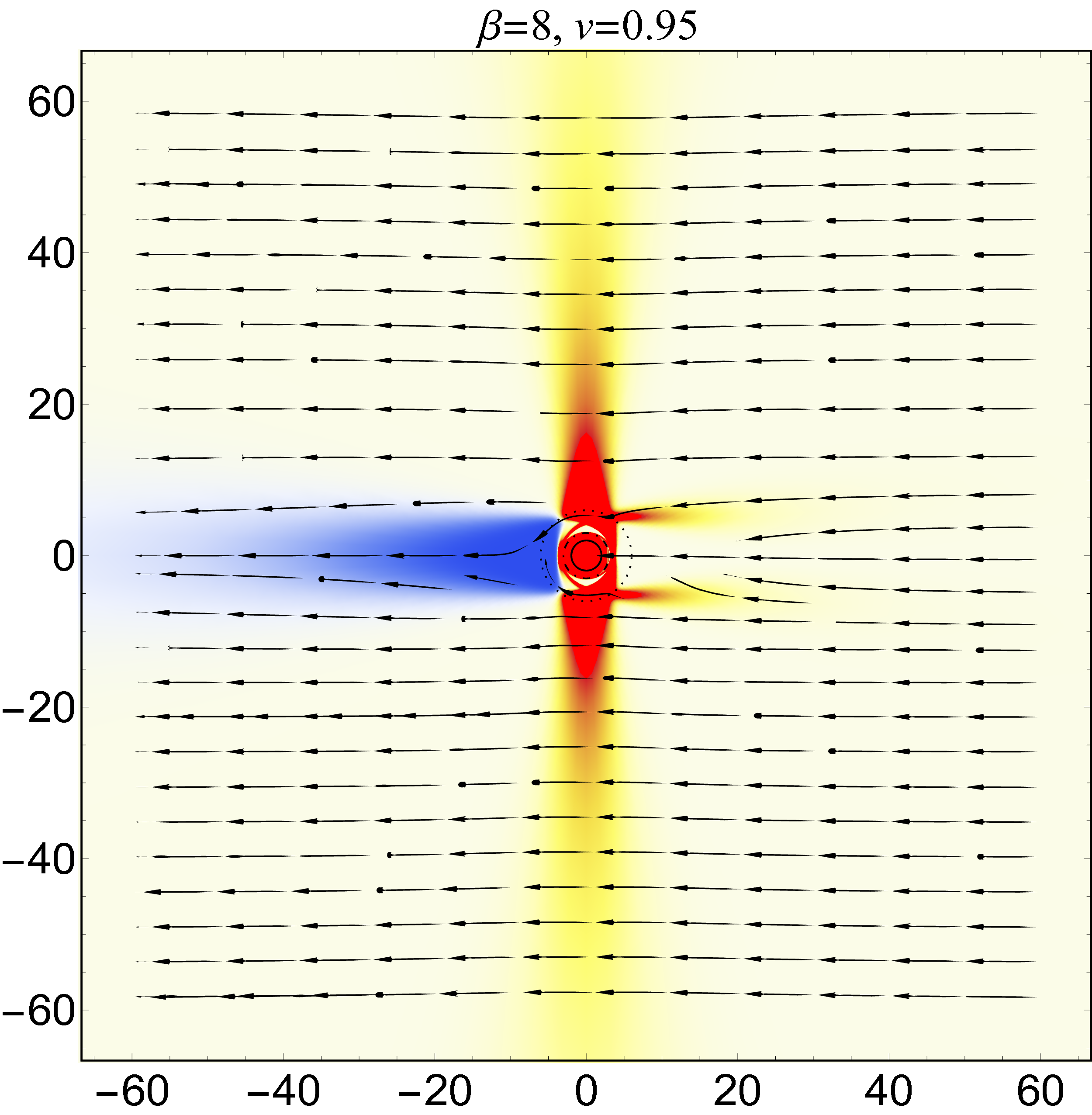}\\
    \caption{\label{fig:Jn_far}
    The dependence of the flow on the black hole velocity $v$ and the parameter $\beta$. The upper row corresponds to $\beta = 1$; the graphs in the lower row were obtained for $\beta = 8$. The columns correspond (from left to right) to $v = 0.05$, $0.5$, and $0.95$, respectively. The colors in the graphs depict the particle density ratio $n/n_\infty$. The vector field $(J^y,J^z)$ is depicted with arrows, similarly to Fig.\ \ref{fig:AbsvsScattvsTotal}.
    }
    
\end{figure*}

\begin{figure*}
    \centering
    \includegraphics[width=0.666\columnwidth]{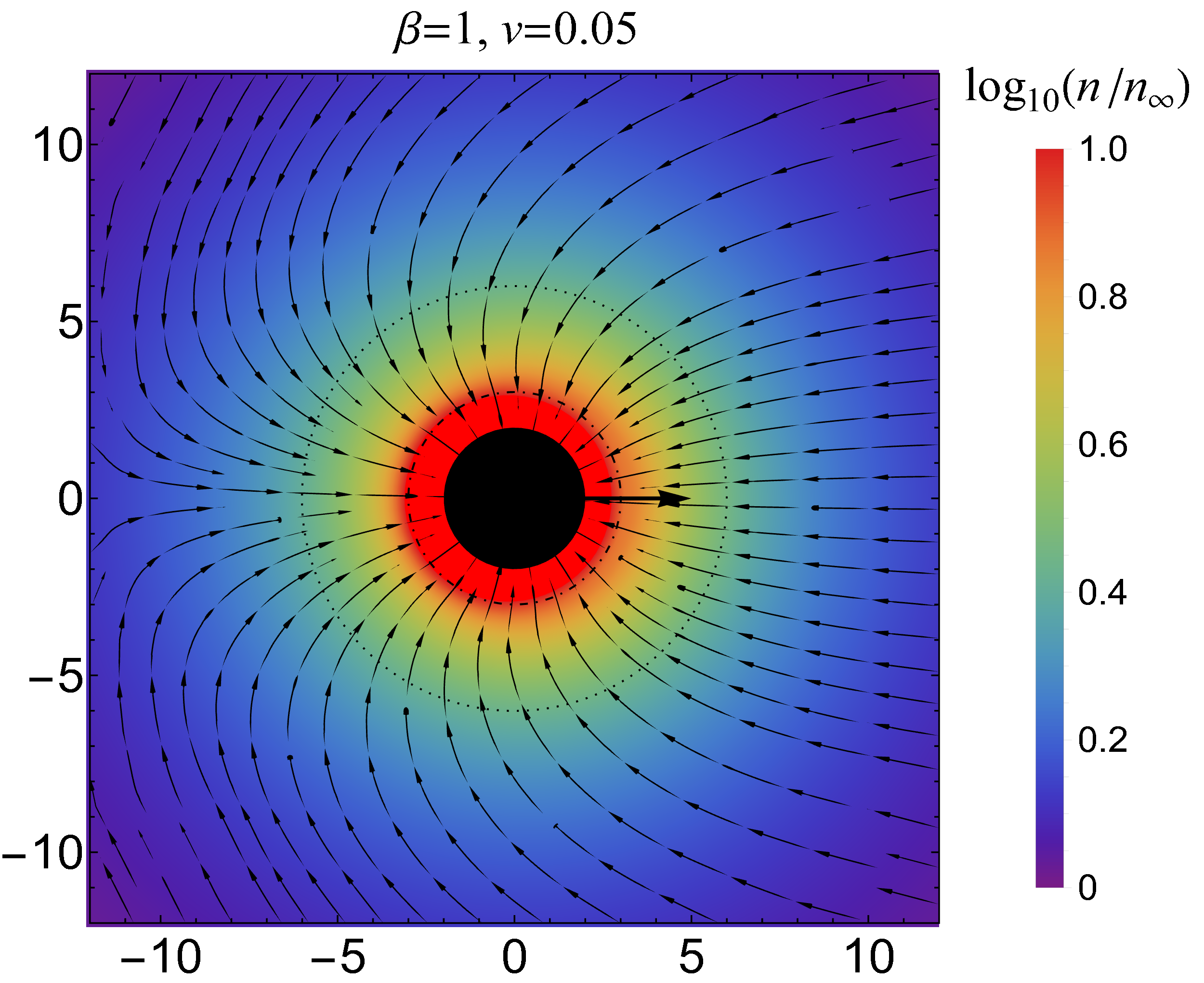}~\includegraphics[width=0.666\columnwidth]{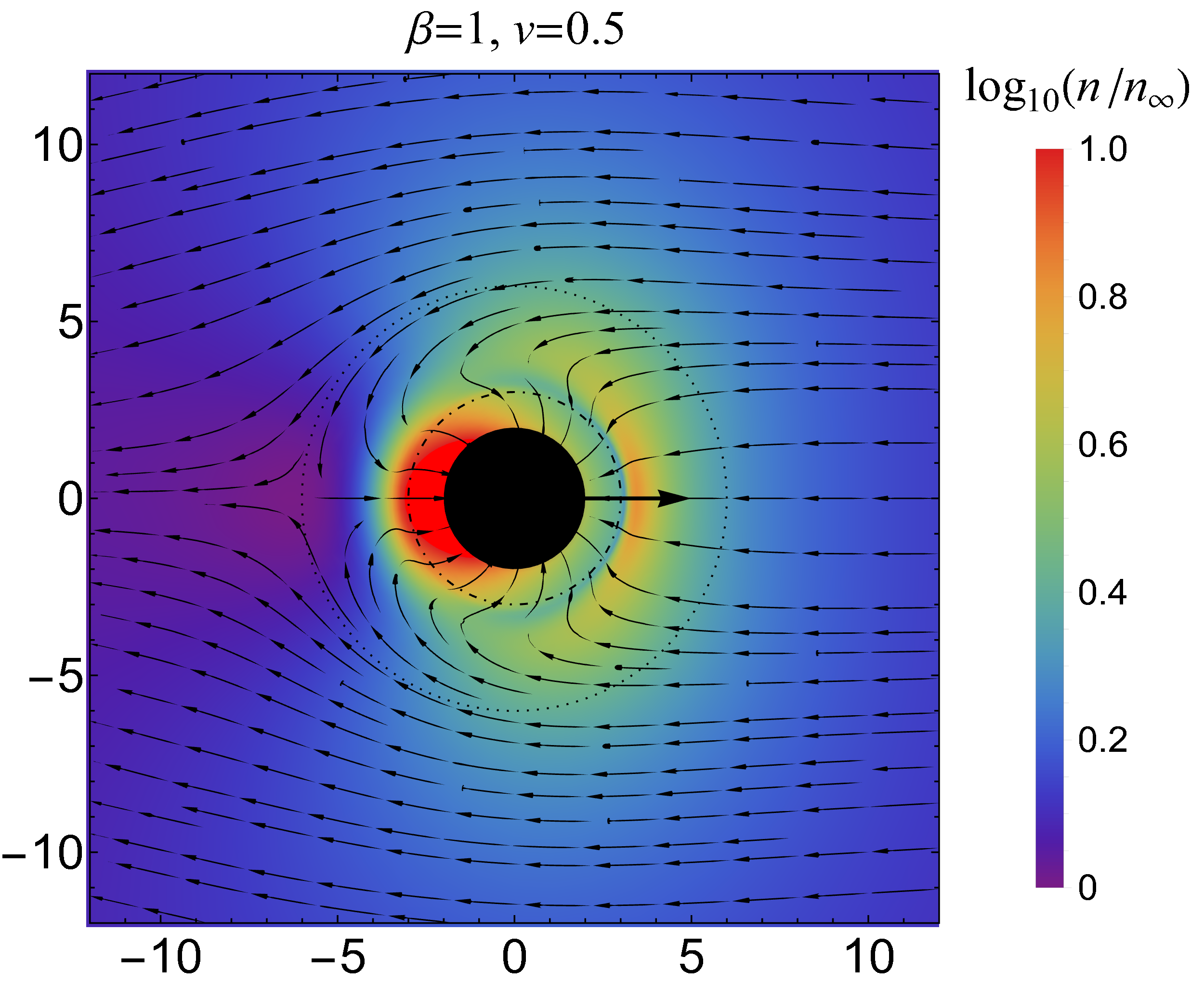}~\includegraphics[width=0.666\columnwidth]{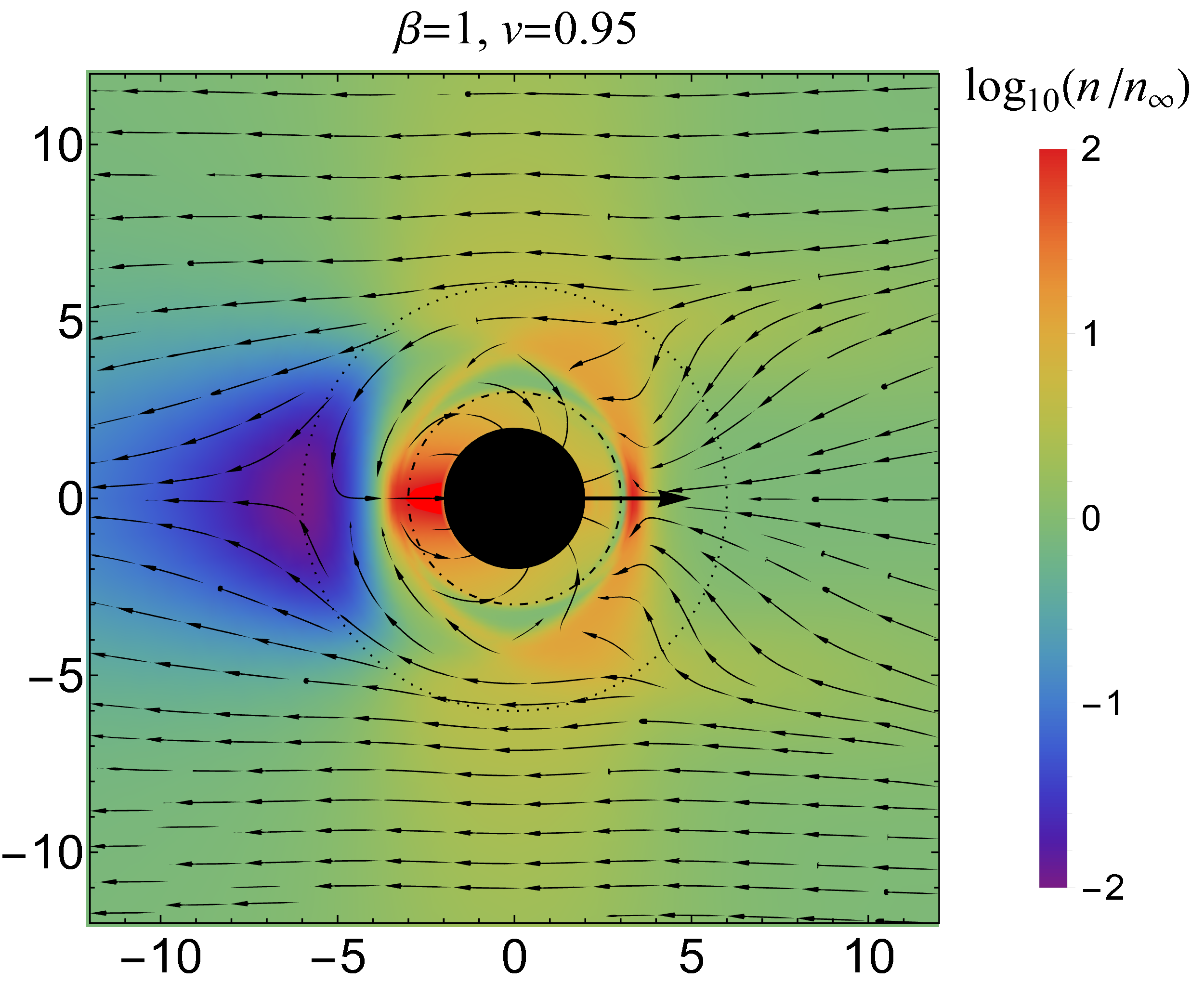}\\
    \includegraphics[width=0.666\columnwidth]{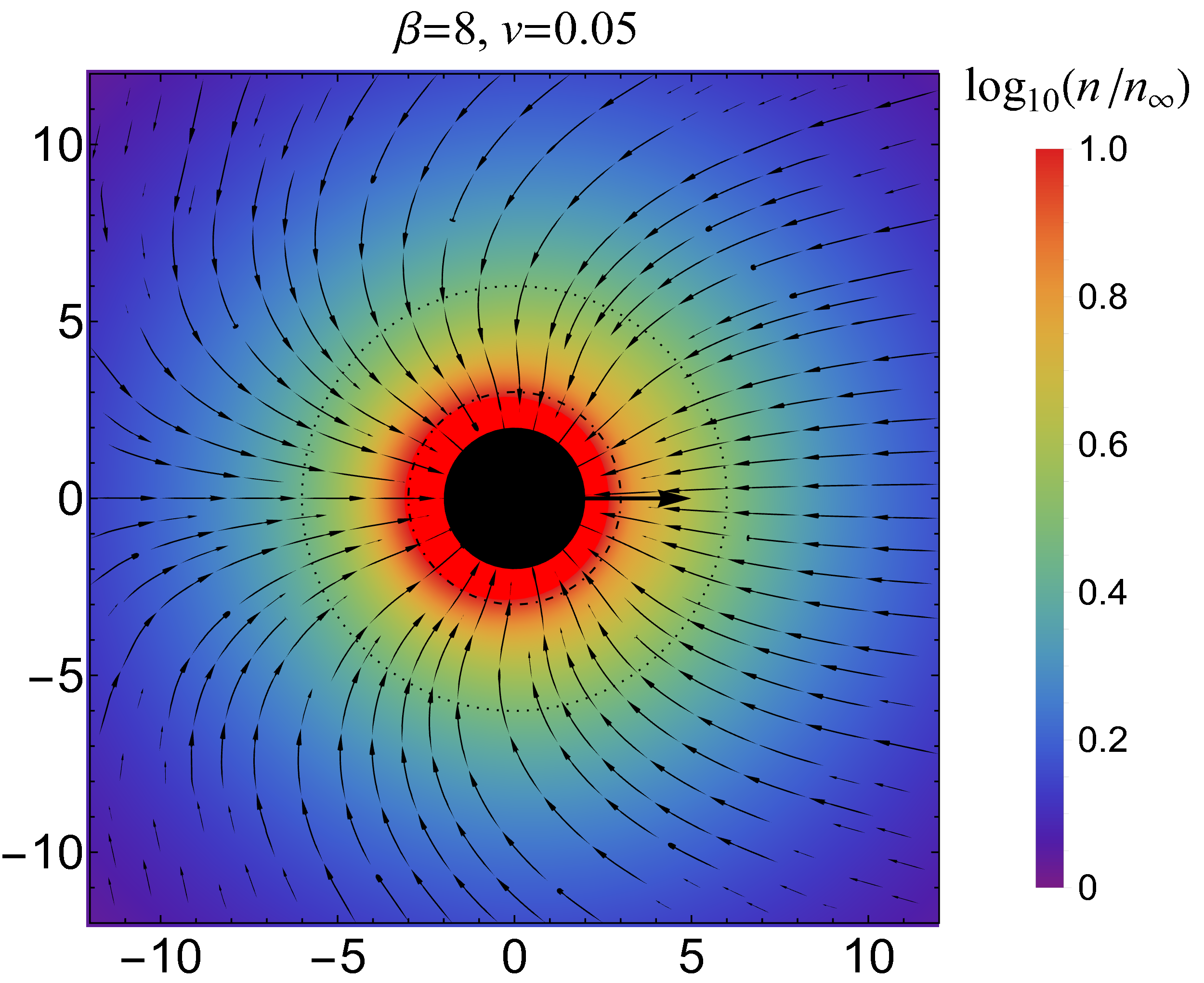}~\includegraphics[width=0.666\columnwidth]{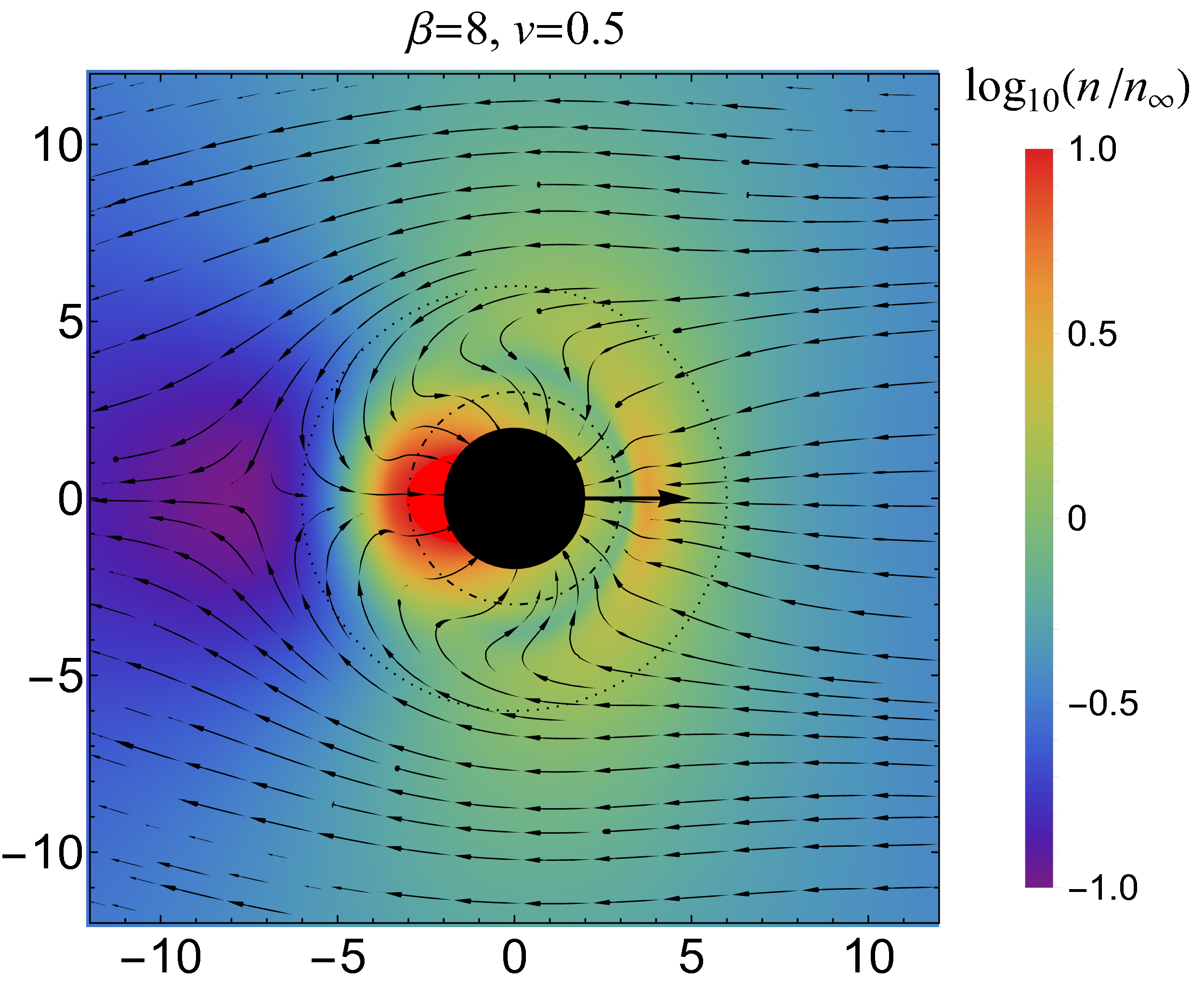}~\includegraphics[width=0.666\columnwidth]{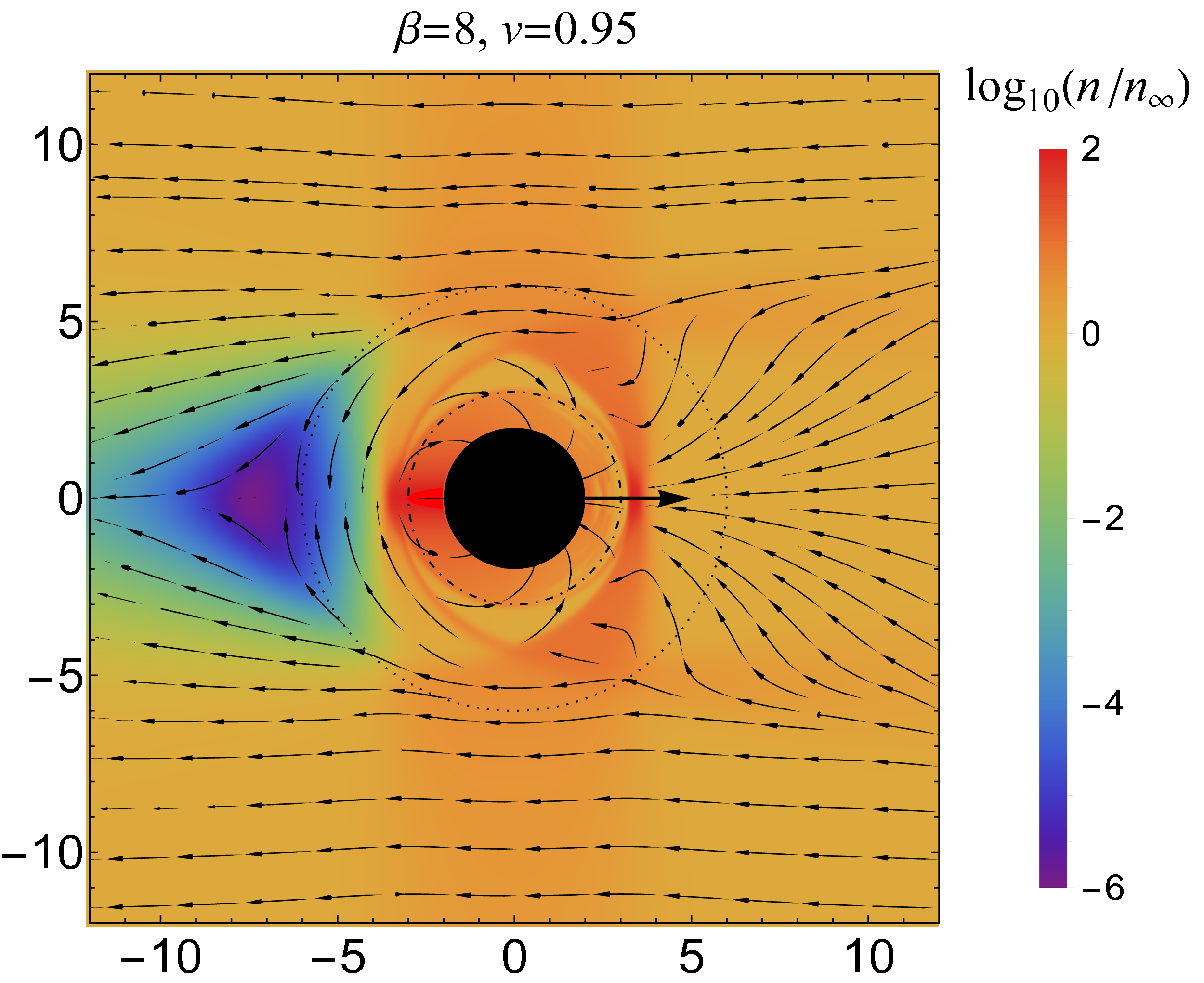}\\
    \caption{\label{fig:Jn_zoom} Same as in Fig.\ \ref{fig:Jn_far}, except for the density scale, which is now logarithmic. The graphs show the morphology of the flow in the vicinity of the black hole. The region inside the horizon is marked in black, although the solution was computed up to $\xi = 1$. Two additional circles with radii $\xi = 3$ and $\xi = 6$ mark the locations of the photon sphere and the innermost stable circular orbit.}
\end{figure*}

\begin{figure}
    \centering
    \includegraphics[width=\columnwidth]{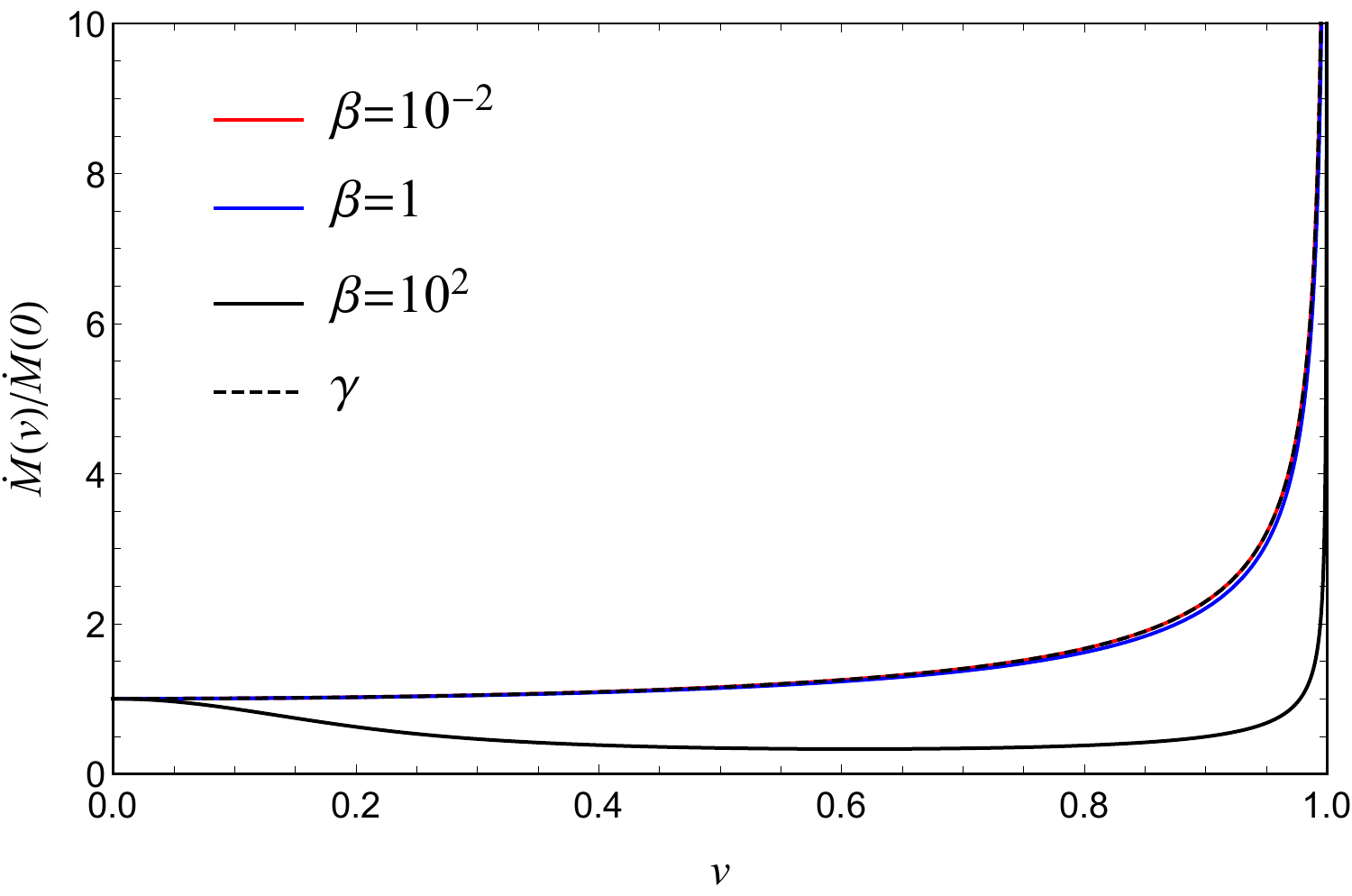}
    \caption{\label{fig:MdotInf1} The ratio $\dot M(v)/\dot M(0)$ of the accretion rate given by Eq.\ \eqref{Mdot_infty} for the moving black hole to the accretion rate corresponding to $v = 0$. For nonrelativistic particles with sufficiently large $\beta > 4.844$ (this threshold value has been determined numerically) the accretion becomes suppressed for moderate black hole speeds $v$. For $v \simeq 1$, the ratio $\dot M(v)/\dot M(0)$ is proportional to the value of the Lorentz factor $\gamma$ for all values of $\beta$ (see also Fig.\ \ref{fig:MdotInf2}).}
\end{figure}

\begin{figure}
    \centering
    \includegraphics[width=\columnwidth]{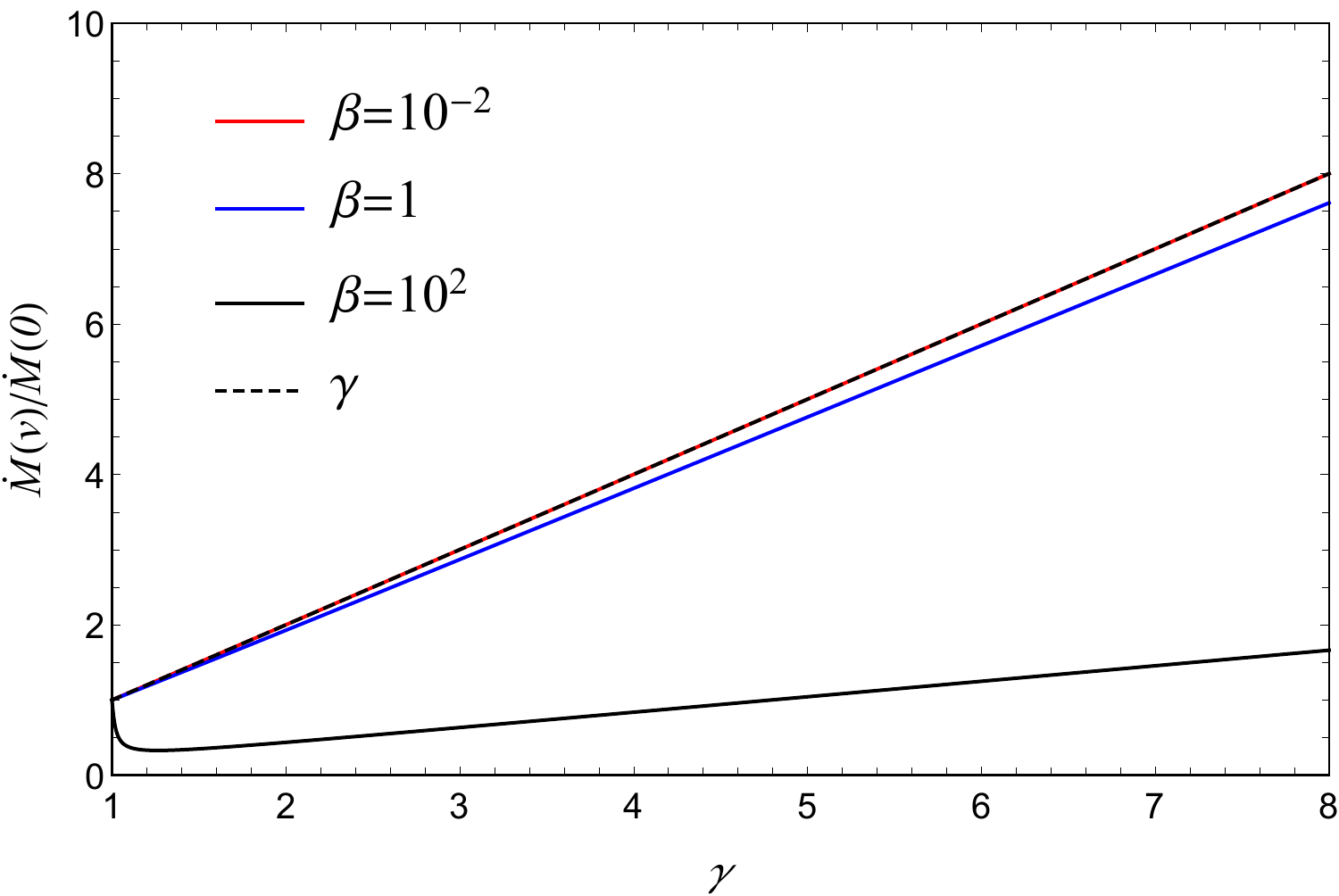}
    \caption{\label{fig:MdotInf2} Same as in Fig.~\ref{fig:MdotInf1}. Instead of the velocity, the abscissa shows the Lorentz factor $\gamma$, to clarify a nearly linear behavior for ultra-relativistic black hole velocities. For sufficiently small $\beta \ll 1$ (red), the ratio $\dot M(v)/\dot M(0)$ is indistinguishable  from $\gamma$ (dashed).}    
\end{figure}

\begin{figure}
    \centering
    \includegraphics[width=\columnwidth]{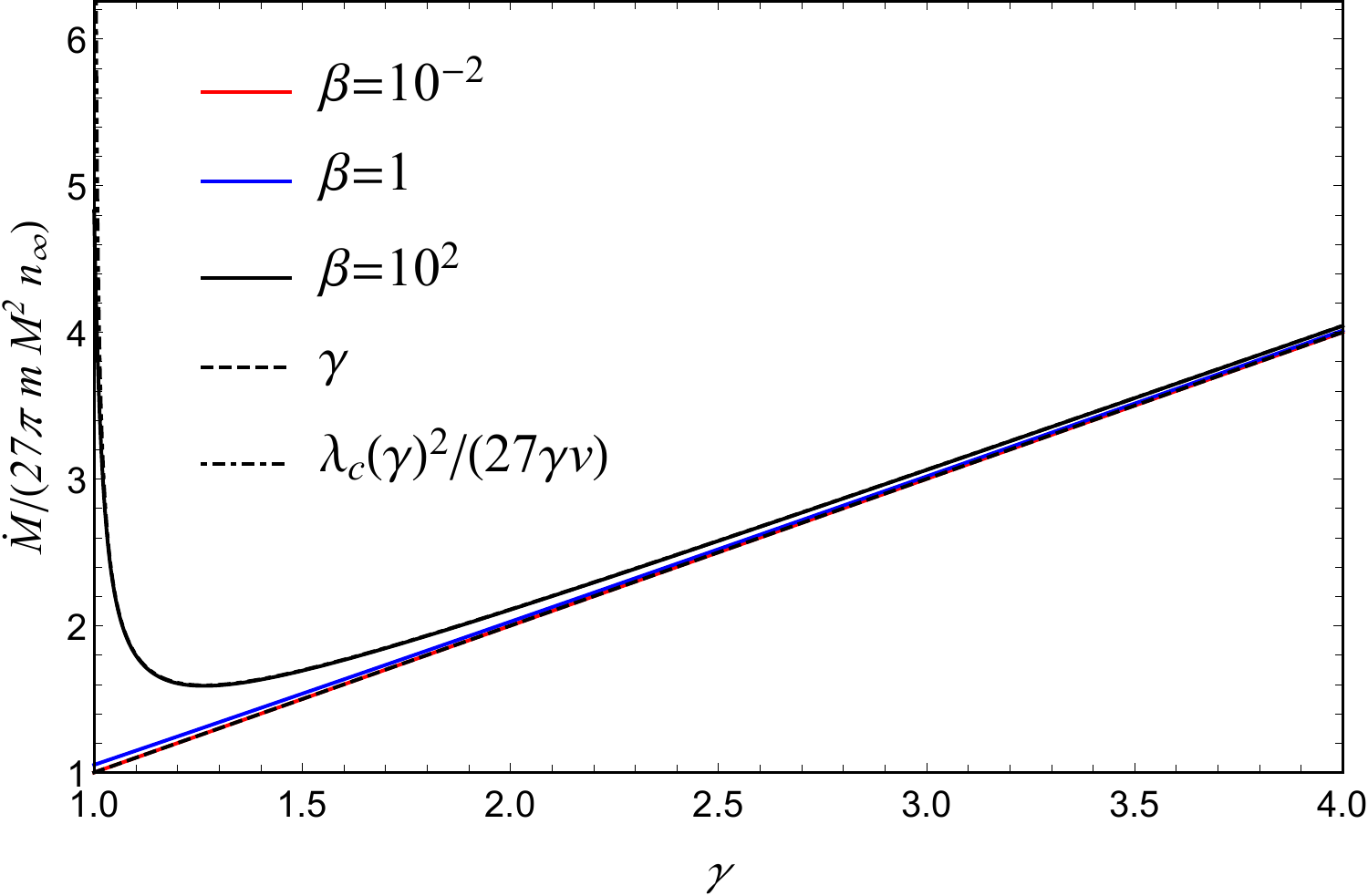}
    \caption{\label{fig:MdotInf3} The ratio $\dot M/(27 \pi m M^2 n_\infty)$ versus $\gamma$. The Lorentz factor provides both the lower estimate and the high-velocity limit. The upper limit for the accretion rate is given by Eq.\ (\ref{lowtemperaturelimit}).}  
\end{figure}

A numerical computation of the integrals representing the components of the particle current density $J_\mu$, derived in the preceding sections, is time consuming and not exactly straightforward, but it is feasible with computer systems such as Wolfram Mathematica \cite{wolfram}. They are all relatively simple, double integrals of elementary functions, with the exception of $X(\xi,\varepsilon,\lambda)$, which is given in terms of elliptic functions (and which is responsible for the main computational cost in our numerical implementation). On the other hand, care should be taken in order to evaluate these integrals to a satisfactory accuracy, as shown by a necessity of replacing the expression (\ref{lambdacorig}) for $\lambda_c(\varepsilon)$ with its more stable version (\ref{limit1}). In this section we show the results of our computations of the particle current density $J_\mu$ and the particle density $n$ for a sample of solutions.

Figures \ref{fig:n} and \ref{fig:AbsvsScattvsTotal} illustrate a sample morphology of the flow, obtained for $v = 0.5$ and $\beta = 8$. In Fig.\ \ref{fig:n} we plot radial and angular profiles of the particle density $n$. As expected, the particle density tends to $n_\infty$, as $\xi \to \infty$ (in all angular directions). Close to the black hole, the structure of the flow becomes quite intricate. There is a local maximum of the particle density in front of the black hole (for $\theta = 0$), just outside the photon sphere, and a minimum behind the black hole (for $\theta = \pi$). The graph of the angular density profiles (Fig.\ \ref{fig:n}, lower panel) depicts also an increase of the particle density in the direction perpendicular to the symmetry axis (i.e., to the direction of the black hole motion).

For visualization purposes we define the Cartesian components of the particle density current 
\begin{eqnarray*}
J^x & = & \left( \frac{J_\theta}{r} \cos{\theta} + J^r \sin{\theta} \right) \cos \varphi \\
J^y & = &\left( \frac{J_\theta}{r} \cos{\theta} + J^r \sin{\theta} \right) \sin \varphi, \\
J^z & = & -\frac{J_\theta}{r} \sin{\theta} + J^r \cos{\theta}.
\end{eqnarray*}
In Figs.\ \ref{fig:AbsvsScattvsTotal}--\ref{fig:Jn_zoom} we plot the components $J^y$ and $J^z$ at the plane $x = 0$, so that $J^x \equiv 0$. Figure \ref{fig:AbsvsScattvsTotal} depicts the structure of the flow, with an explicit division into the absorbed part $J^\mu_\mathrm{(abs)}$, the scattered part $J^\mu_\mathrm{(scat)}$, and the total current $J^\mu = J^\mu_\mathrm{(abs)} + J^\mu_\mathrm{(scat)}$.
The colors in Fig.\ \ref{fig:AbsvsScattvsTotal} depict the length of the spatial part of the particle density current
\[ |\vec{J}| = \sqrt{g_{ij} J^i J^j} =  \sqrt{ \left( 1+\frac{2M}{r} \right) \left(J^r \right)^2 +\left(\frac{J_\theta}{r}\right)^2}, \]
normalized by the particle density $n$. Far from the black hole (upper panels), the absorbed flow is roughly spherically symmetric, while the scattered one represents a uniform motion along the symmetry axis. For the case depicted in Fig.\ \ref{fig:AbsvsScattvsTotal}, the uniform motion prevails in the total current (upper right panel), but this does not need to be the case---note that for $v = 0$ (spherically symmetric flow), we have $J^r_\mathrm{(scat)} = 0$. For sufficiently small velocities $v$, we would expect a region around the black hole, in which the accretion occurs roughly spherically symmetrical; at the same time at far distances from the black hole the uniform motion with a nonzero velocity $v$ should dominate the radial velocity component, which vanishes asymptotically. In the examples shown in Figs.\ \ref{fig:Jn_far} and \ref{fig:Jn_zoom} for $v = 0.05$, the approximately spherically symmetric region has a radius smaller than $\xi \approx 10$. This roughly spherically symmetric region seems to disappear for black hole velocities larger than $v \approx 0.1$.

Among other structures present in Fig.\ \ref{fig:AbsvsScattvsTotal}, we note the importance of the photon sphere at $\xi = 3$ (dot-dashed circle) and, to a less extent, the innermost stable circular orbit with the radius $\xi = 6$ (dotted circle). An inspection of the flow in the vicinity of the black hole (lower row) reveals the existence of a stagnation point behind the black hole. This is a common feature observed in all our models, and also in hydrodynamical solutions of the relativistic Bondi-Hoyle-Lyttleton accretion.

The dependency of the solutions on $v$ and $\beta$ is shown in Figs. \ref{fig:Jn_far} and \ref{fig:Jn_zoom}.  The morphology of the flow seems to be affected much stronger by the black hole velocity $v$ than by the parameter $\beta$. As expected, for small black hole velocities $v$ the accretion occurs more or less isotropically (Fig.\ \ref{fig:Jn_zoom}, left column). On the other hand, for relativistic black hole speeds, anisotropic features, signalled already in Fig.\ \ref{fig:n}, become dominant. In the large scale, we observe mainly the increase of the particle number density in the direction perpendicular to the symmetry axis, which becomes thinner and thinner with an increasing velocity $v$. This seems to be consistent with a hand-waved explanation in terms of the relativistic Lorentz contraction. There is also a long tail of a low-density material behind the black hole. The matter gets compressed in front of the photon sphere, but also just behind the black hole, where it is accreted through the horizon.

Another peculiar feature occurring for ultrarelativistic black hole speeds (Figs.\ \ref{fig:Jn_far} and \ref{fig:Jn_zoom}, right column) is the presence of a high-density tubelike region in front of the black hole. It is roughly parallel to the symmetry axis and forms a kind of a funnel, dividing the stream accreted directly by the black hole from the stream ``passing by.''

We would also like to point the reader's attention to huge differences in the value of the particle density at its minimum close to the stagnation point. For $\beta = 8$ and $v = 0.95$ the particle density $n$ at this minimum is nearly six orders of magnitude smaller than the asymptotic value $n_\infty$. For $\beta = 1$ and $v = 0.95$, the ratio of $n/n_\infty$ in this rarefaction region is of the order $n/n_\infty \approx 10^{-2}$. This appears to be the only quantity rapidly changing with $\beta$.

Figures \ref{fig:MdotInf1}--\ref{fig:MdotInf3} show the dependence of the accretion rate $\dot M$ on the black hole velocity $v$ and the parameter $\beta$. In Fig.\ \ref{fig:MdotInf1} we plot the ratio of the accretion rate $\dot M = \dot M(v)$ to its value corresponding to the case with $v = 0$, denoted as $\dot M(0)$. For nonrelativistic particles ($\beta > 4.844$), the ratio $\dot M(v)/\dot M(0)$ is not monotonic with $v$, and it drops below 1 for moderate black hole velocities $v$.  For ultrarelativistic black hole velocities, the ratio $\dot M(v)/\dot M(0)$ turns out to be proportional to the Lorentz factor $\gamma$. The same, nearly linear dependence on $\gamma$ is also observed for small values of the parameter $\beta$ and the entire range of $v$ (see Fig.\ \ref{fig:MdotInf2}). This confirms the limit given by Eq.\ (\ref{limitbeta0}), derived in Appendix \ref{appendixb}. Figure \ref{fig:MdotInf3} illustrates this fact in yet another way, by plotting the ratio $\dot M/(27 \pi M^2 m n_\infty)$ versus $\gamma$.

\section{Conclusions}
\label{sec:conclusions}

Using the formalism developed by Rioseco and Sarbach in \cite{Olivier}, we have been able to derive an exact solution representing stationary accretion of the relativistic Vlasov gas onto a moving Schwarzschild black hole. In general, the obtained accretion rate is not a monotonic function of the black hole velocity, although in the limit of $\beta \to 0$ (hot gases) we recover the situation known from the relativistic accretion of ultrahard fluids---the accretion rate is directly proportional to the Lorentz factor associated with the black hole velocity. In the low-temperature limit, the behavior of the Vlasov model is similar to the result of the ballistic approximation used in \cite{tajeda}---the accretion rate attains a local minimum at a finite, nonzero black hole velocity.

As usual for the relativistic accretion onto a moving black hole, we describe the velocity field in the frame associated with the black hole. For highly relativistic black hole speeds the velocity of the gas departures from the uniform asymptotic distribution only in a narrow zone, parallel to the symmetry axis. In all cases there is a stagnation point behind the black hole in which the velocity of the gas with respect to the black hole vanishes. It is also roughly correlated with a minimum of the particle number density. Remarkably, for highly relativistic black hole speeds the density in this minimum can be lower than its asymptotic value by several orders of magnitude. A substantial increase of the density is observed in front of the black hole, still outside the photon sphere, and behind the black hole, in a close vicinity of the horizon.

Our analysis is limited to stationary configurations of a Vlasov gas of collisionless and nondegenerate particles. Obvious possible generalizations include taking into account Fermi-Dirac and Bose-Einstein statistics, scattering terms between particles, dynamics of the flow, the existence of particles on bounded orbits. The latter would become important, if the collisions between the particles or self-gravity of the gas were taken into account. Also a natural (but probably difficult) generalization would consist of considering a Kerr black hole instead of the Schwarzschild spacetime. We believe that all these generalizations are technically feasible to some extent.

\begin{acknowledgments}
P.\ M.\ was partially supported by the Polish National Science Centre Grant No.\ 2017/26/A/ST2/00530.
\end{acknowledgments}

\appendix

\section{Derivation of the expression for $Q^3$}
\label{appendixa}

In this appendix we derive expressions (\ref{Q3}) and (\ref{X}) for $Q^3$ in the Schwarzschild spacetime. Recall that $Q^3$ is defined as [Eq.\ (\ref{q3})]
\[ Q^3 = - l \int_\Gamma \frac{dr}{r^2 \left( -g^{tr} E + g^{rr} p_r \right)} + l \int_\Gamma \frac{d \theta}{p_\theta}. \]
The second integral can be evaluated in terms of elementary functions. Recalling that $p_\theta$ is given by Eq.\ (\ref{pthetaformula}), we get
\[ l \int_\Gamma \frac{d\theta}{p_\theta} = - \epsilon_\theta \arctan \left( \frac{l \cot \theta}{\sqrt{l^2 - \frac{l_z^2}{\sin^2 \theta}}} \right) + \mathrm{const}. \]
The integral
\[ - l \int_\Gamma \frac{dr}{r^2 \left( -g^{tr} E + g^{rr} p_r \right)} \]
is more problematic. A direct calculation making use of Eq.\ (\ref{pixi}) yields
\begin{eqnarray*}
\lefteqn{- l \int_\Gamma \frac{dr}{r^2 \left( -g^{tr} E + g^{rr} p_r \right)} }\\
&&=  - \lambda \int_\Gamma \frac{d \xi}{\xi^2 \left( - g^{tr} \varepsilon + g^{rr} \pi_\xi \right)}  \\
&& = - \lambda \epsilon_r \int \frac{d\xi}{\xi^2 \sqrt{\varepsilon^2 - \left(1 - \frac{2}{\xi}\right)\left(1 + \frac{\lambda^2}{\xi^2}\right)}},
\end{eqnarray*}
where we have assumed that $\epsilon_r$ is fixed along $\Gamma$. We fix the integration constant by introducing
\begin{equation}
\label{Xappendix1}
X(\xi,\varepsilon,\lambda) = \lambda \int_\xi^\infty \frac{d \xi^\prime}{{\xi^\prime}^2 \sqrt{\varepsilon^2 - \left(1 - \frac{2}{\xi^\prime} \right) \left( 1 + \frac{\lambda^2}{{\xi^\prime}^2} \right)}} \end{equation}
[Eq.\ (\ref{X})] and setting
\[ Q^3 = \epsilon_r X(\xi,\varepsilon,\lambda) - \epsilon_r \frac{\pi}{2} - \epsilon_\theta \arctan \left( \frac{\lambda \cot \theta}{\sqrt{\lambda^2 - \frac{\lambda_z^2}{\sin^2 \theta}}} \right). \]
[Eq.\ (\ref{Q3})]. Asymptotically (for $\xi \to \infty$), the radial terms tend to $- \epsilon_r \pi/2$, which agrees with the asymptotic limit of Eq.\ (\ref{q3mink}).

The integral (\ref{Xappendix1}) can be evaluated by substituting $w = a \left( \frac{1}{\xi} - \frac{1}{6} \right)$, where $a > 0$ is an arbitrary constant. This yields
\begin{equation}
\label{Xappendix2}
X = - \sqrt{2a} \int_{\frac{a}{\xi} - \frac{a}{6}}^{-\frac{a}{6}} \frac{dw}{\sqrt{4 w^3 - g_2 w - g_3}},
\end{equation} 
where
\begin{eqnarray*}
g_2 & = & - \frac{2}{\lambda^2} a^2 \left( 2 - \frac{\lambda^2}{6} \right), \\
g_3 & = & -\frac{2}{\lambda^2} a^3 \left( \varepsilon^2 - \frac{2}{3} - \frac{\lambda^2}{54} \right).
\end{eqnarray*}
The integral (\ref{Xappendix2}) can be expressed in terms of an inverse of a restriction of the Weierstrass elliptic function $\wp$ by recalling the standard integral formula for the Weierstrass function
\[ z = \int_{\wp(z;g_2,g_3)}^\infty \frac{dw}{\sqrt{4 w^3 - g_2 w - g_3}}. \]

In numerical applications inverting the Weierstrass function $\wp$ can be inconvenient because of a necessity of choosing its appropriate restrictions. In practice, we use the representation of the indefinite integral
\[ I(w) = \int \frac{dw}{\sqrt{4 w^3 - g_2 w - g_3}} \]
in the form
\begin{eqnarray}
\label{indefiniteI}
I(w) & = & \frac{2 \left(w-w_3\right) \sqrt{\frac{w-w_1}{w_3-w_1}} \sqrt{\frac{w-w_2}{w_3-w_2}} }{\sqrt{\frac{w-w_3}{w_2-w_3}} \sqrt{4 w^3 - g_2 w - g_3}} \\
&& \times F\left(\arcsin \sqrt{\frac{w_3-w}{w_3-w_2}}|\frac{w_2-w_3}{w_1-w_3}\right) + C, \nonumber
\end{eqnarray}
where $w_1,w_2,w_3$ denote (possibly complex) roots of the polynomial $4 w^3 - g_2 w - g_3 = 0$, and $F(\phi|m)$ is Legendre's elliptic integral of the first kind. The convention for $F(\phi | m)$ used here is
\[ F(\phi | m) = \int_0^\phi \frac{d \theta}{\sqrt{1 - m \sin^2 \theta}}, \]
for $-\pi/2 < \phi < \pi/2$. The integral $X(\xi,\varepsilon,\lambda)$, given by Eq.\ (\ref{Xappendix2}), is then computed as
\[ X = -\sqrt{2a} \left[ I(-a/6) - I(a/\xi - a/6) \right]. \]
In our numerical calculations, the above formula yields a correct, real result, even though the representation of the indefinite integral $I(w)$ given by Eq.\ (\ref{indefiniteI}) is in general complex valued and depends on the ordering of $w_1$, $w_2$, and $w_3$. Since $a$ can be any positive constant, we assume $a = 1$.

\section{Derivation of the limiting expressions for $\dot{M}$}
\label{appendixb}

\begin{figure}
\includegraphics[width=\columnwidth]{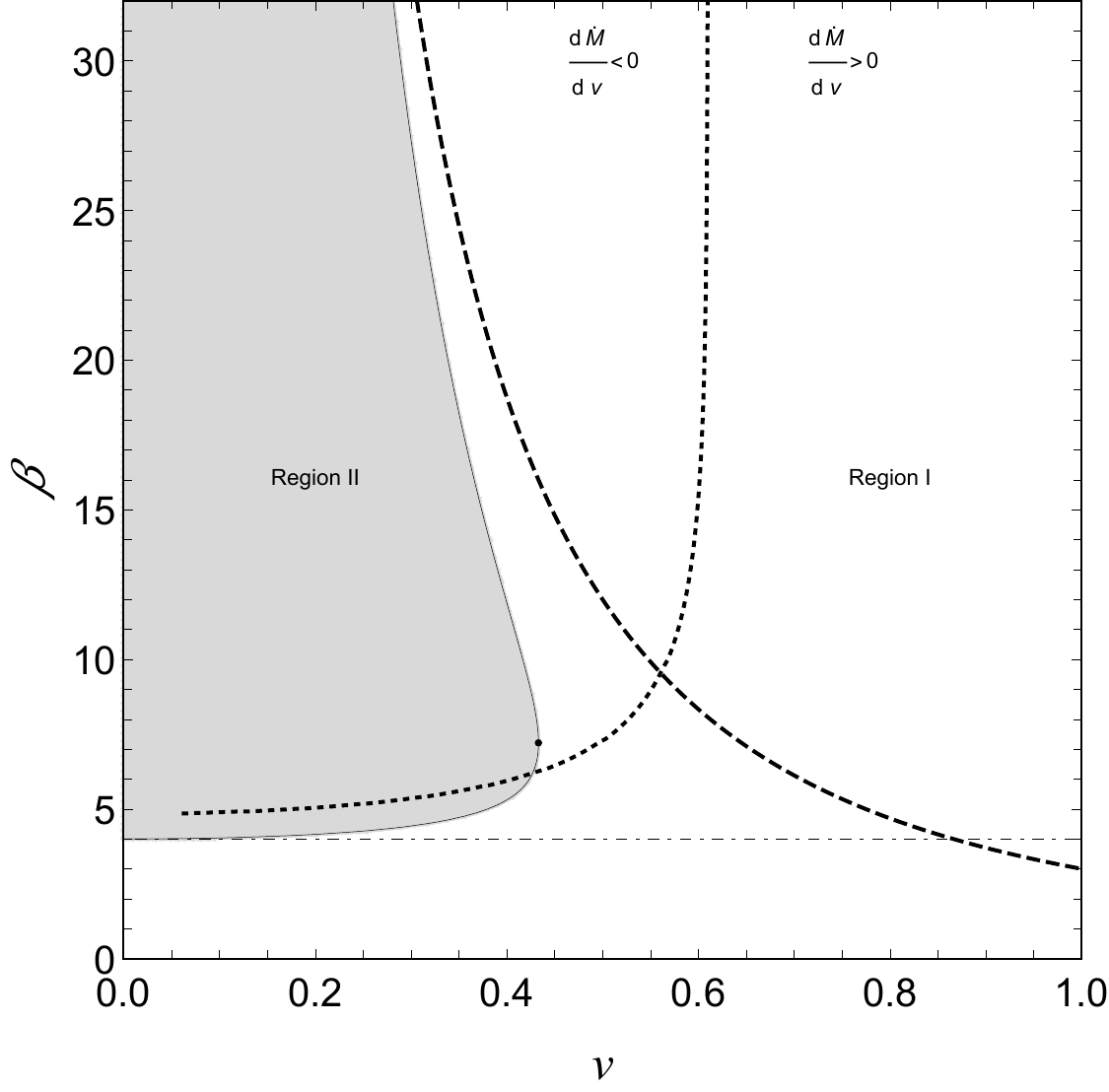}
\caption{
\label{regions_area}
Region I (white) consists of points $(v,\beta)$ for which integrand (\ref{Mdot_integrand_full}) has a local maximum in the range $\varepsilon > 1$. Region II (shaded) consists of points $(v,\beta)$ for which integrand (\ref{Mdot_integrand_full}) is a decreasing function of $\varepsilon$ for $\varepsilon > 1$. Region II is bounded by the line $\beta = 4$ and extends up to the point $(v,\beta) = (\sqrt{3}/4, 2\sqrt{13})$ (black dot). The curve $v^2 = 3/\beta$, i.e., an approximation of the boundary between regions I and II, valid for large values of $\beta$, is depicted with the dashed line. The dotted line shows a numerically determined location of points satisfying the condition $d \dot{M}/dv = 0$. 
}
\end{figure}

In this appendix we compute limiting expressions for the mass accretion rate $\dot M$ given in Sec.\ \ref{sec:jcalculation}. To shorten the notation, we define $\rho_\infty = m n_\infty$.

We start the discussion with the high-temperature limit ($\beta \to 0$). Without loosing generality, we assume $v \ge 0$, so that $\gamma v = \sqrt{\gamma^2 - 1}$, and  we write the integrand appearing in Eq.\ (\ref{Mdot_infty}) as
\begin{eqnarray}
\lefteqn{ e^{-\beta \gamma \varepsilon} \lambda_c(\varepsilon)^2 \frac{\sinh \left( \beta \gamma v \sqrt{\varepsilon^2 - 1} \right)}{{\beta \gamma v \sqrt{\varepsilon^2-1}}}} \nonumber \\
&& = e^{-\beta \gamma \varepsilon} \lambda_c(\varepsilon)^2 \frac{\sinh \left(\beta \sqrt{\gamma^2 - 1} \sqrt{\varepsilon^2-1} \right)}{{\beta \sqrt{\gamma^2 - 1} \sqrt{\varepsilon^2-1}}}.
\label{Mdot_integrand_full}
\end{eqnarray}
The above expression can be bounded from below and from above by noticing that
\begin{equation}
\label{lambdac_approx}
27 \varepsilon^2 -9 -\frac{2}{\varepsilon} \leq \lambda_c(\varepsilon)^2 < 27 \varepsilon ^2 -9 -\frac{1}{\varepsilon ^2}
\end{equation}
and
\[ \varepsilon -1 \le \sqrt{\varepsilon^2 - 1} < \varepsilon \]
for $\varepsilon \ge 1$. Since for positive arguments $x$, the function $\sinh (x)/x$ increases with $x$, we obtain the following upper and lower estimates for $\dot M$:
\begin{eqnarray*}
\lefteqn{\dot M  <  \dot{M}_\mathrm{est}^{+} = \pi M^2 \rho_\infty \frac{\beta}{K_2(\beta)}  \times} \\
&& \int_1^\infty d \varepsilon e^{-\beta \gamma \varepsilon} \left( 27  \varepsilon^2 - 9 - \frac{1}{\varepsilon^2} \right) \frac{\mathrm{sinh} \left( \beta \sqrt{\gamma^2 - 1} \varepsilon \right)}{\beta \sqrt{\gamma^2 - 1} \varepsilon},
\end{eqnarray*}
\begin{eqnarray*}
\lefteqn{\dot M  >  \dot{M}_\mathrm{est}^{-} = \pi M^2 \rho_\infty \frac{\beta}{K_2(\beta)} \times } \\
&&  \int_1^\infty d \varepsilon e^{-\beta \gamma \varepsilon} \left( 27  \varepsilon^2 - 9 - \frac{2}{\varepsilon} \right) \frac{ \mathrm{sinh} \left[ \beta \sqrt{\gamma^2 - 1} (\varepsilon-1) \right]}{\beta \sqrt{\gamma^2 - 1} (\varepsilon-1)}. 
\end{eqnarray*}
The integrals appearing in the above expressions for $\dot{M}_\mathrm{est}^{\pm}$ can be computed analytically, say with Wolfram Mathematica \cite{wolfram}. The resulting expressions are lengthy, but simple. In both cases, one can compute the limit of $\beta \to 0$, which reads
\[ \lim_{\beta \to 0} \dot{M}_\mathrm{est}^{\pm} = 27 \pi M^2 \rho_\infty \gamma.  \]
Since $\dot{M}_\mathrm{est}^{-} < \dot M < \dot{M}_\mathrm{est}^{+}$, one obtains Eq.\ (\ref{limitbeta0}).

Another possibility to derive the limit of $\dot M$ as $\beta \to 0$, only slightly less straightforward, is to substitute $y = \beta \gamma \varepsilon$ in the integral in Eq.\ (\ref{Mdot_infty}) and use Lebesgue's dominated convergence theorem.

Using Lebesgue's dominated convergence theorem, Rioseco and Sarbach \cite{Olivier} provided the $\beta \to \infty$ limit of $\dot M$ in the $v = 0$ case. It reads
\[ \lim_{\beta \to \infty} \frac{\dot M(v = 0)}{\sqrt{\beta}} = 16 \sqrt{2 \pi} M^2 \rho_\infty. \]

A direct application of the above strategies fails for the limit $\beta \to \infty$ and $v > 0$, however even in this case we can provide an approximate expression for $\dot M$. Its derivation relies on the analysis of the location of the maximum in the integrand (\ref{Mdot_integrand_full}) in the range $\varepsilon > 1$. In general, the integrand (\ref{Mdot_integrand_full}) can behave in two distinct ways. It can either be a decreasing function of $\varepsilon$ for the entire integration range $\varepsilon > 1$, or it can have a local maximum for some $\varepsilon > 1$. We depict these two possibilities in the parameter space of $(v,\beta)$ in Fig.\ \ref{regions_area}. Region I (white) consists of all points $(v,\beta)$ for which integrand (\ref{Mdot_integrand_full}) has a local maximum for $\varepsilon > 1$. Region II (shaded) consists of all values $(v,\beta)$ for which integrand (\ref{Mdot_integrand_full}) is a decreasing function of $\varepsilon$ for $\varepsilon > 1$.  The boundary between these two regions can be characterized in a standard way by computing the derivative of (\ref{Mdot_integrand_full}) with respect to $\varepsilon$, setting $\varepsilon = 1$, and equating the result to zero. This yields the condition for region II in the form
\[ 12 + v^2 \beta^2 \gamma^2 < 3 \beta \gamma. \]
For large values of $\beta$ the boundary between the two regions can be approximated by $v^2 = 3/\beta$ (dashed line in Fig.\ \ref{regions_area}). This means that for $\beta \to \infty$ we are effectively in region I for all $v > 0$. On the other hand, the points with $v = 0$ and large values of $\beta$ always remain in region II. Consequently, the case with $v = 0$ requires a separate treatment. We will now proceed with constructing an approximation for the mass accretion rate $\dot M$ valid for large values of $\beta$ in region I.

We start this construction by dropping the rapidly decaying term $-\exp(\beta \gamma v \sqrt{\varepsilon^2-1})/2$ in the definition of sinh function in integrand (\ref{Mdot_integrand_full}). The accretion rate can be then approximated as
\begin{equation}
\dot M \approx \dot{M}_\mathrm{I} = \frac{\pi M^2 \rho_\infty \beta}{K_2(\beta)}\int_1^\infty d\varepsilon \; \lambda_c^2\; \frac{e^{\beta \gamma v \sqrt{\varepsilon^2-1} - \beta \gamma \varepsilon}}{2 \beta \gamma v \sqrt{\varepsilon^2-1}}.
\end{equation}
In the next step we change the integration variable: $\varepsilon = \cosh{t}$,
$\sqrt{\varepsilon^2-1} = \sinh{t}$, $d\varepsilon = dt \sqrt{\varepsilon^2-1}$ and obtain
\[ \dot{M}_\text{I} =  \frac{\pi M^2 \rho_\infty }{2 K_2(\beta) \gamma v}\int_0^\infty dt \; \lambda_c^2 \; e^{\beta \gamma v \sinh{t} - \beta \gamma \cosh{t}}. \]
The expression in the exponent has a maximum at $t = \mathrm{artanh}(v)$, i.e., for $\varepsilon = \gamma$. Note that this approximate location of the integrand maximum could be also useful in numerical integration, providing a hint for algorithms which, if unguided, might fail to locate the maximum. The Taylor expansion of the term $\gamma v \sinh{t} - \gamma \cosh{t}$ at this maximum reads
\[ \gamma v \sinh{t} - \gamma \cosh{t} \approx -1 - \frac{1}{2} [t - \mathrm{artanh}(v)]^2 + \ldots \]
The integral in $\dot M_\mathrm{I}$ can be cast into the Gaussian form, roughly following the so-called method of steepest descent (Laplace's method). This gives
\begin{equation}
\label{inf_limit_step_3}
\dot{M}_\text{I} \approx \frac{\pi M^2 \rho_\infty }{2 K_2(\beta) \gamma v} e^{-\beta}  \lambda_c(\gamma)^2  \int_{-\infty}^\infty dt 
  e^{- \frac{1}{2} \beta [t - \mathrm{artanh}(v)]^2}.
\end{equation}
The above Gaussian integral reads
$\sqrt{2 \pi/\beta}$, and the right-hand side of Eq.\ (\ref{inf_limit_step_3}) becomes
\begin{equation}
\pi M^2 \rho_\infty \lambda_c(\gamma)^2 \frac{\sqrt{2\pi} e^{-\beta} }{\sqrt{ \beta} K_2(\beta)\gamma v}.
\end{equation}
Taking the limit $\beta \to \infty$ we finally obtain
\begin{equation}
\dot M_\mathrm{I} \approx \pi  \frac{\lambda_c(\gamma)^2}{\gamma v} \; M^2 \rho_\infty.
\end{equation}

In region II the function $\mathrm{sinh}(x)/x$ with $x = \beta \gamma v \sqrt{\varepsilon^2 - 1}$ in Eq.\ (\ref{Mdot_integrand_full}) can be approximated
by the Maclaurin series
\[ \frac{\sinh{x}}{x} \approx 1 + \frac{1}{6} x^2 + \ldots \]
Using the lower estimate for $\lambda_c(\varepsilon)$ from Eq.\ (\ref{lambdac_approx}), or simply $\lambda_c(1)=4$, one can evaluate the integral
\begin{eqnarray*}
\dot M & \approx & \dot{M}_\mathrm{II} = \frac{\pi M^2 \rho_\infty \beta}{K_2(\beta)} \\
&& \times \int_1^\infty d\varepsilon e^{-\beta \gamma \varepsilon} \lambda_c^2
\left[ 1+ \frac{1}{6} \beta^2 (\gamma^2-1)(\varepsilon^2-1) \right] 
\end{eqnarray*}
analytically. For nonrelativistic black hole velocities $v^2 \ll 3/\beta$ and $\beta \gg 4$ we obtain
\[ \dot{M}_\mathrm{II} \approx \dot{M}(v=0) \times \left(1-\frac{1}{6} \beta v^2 \right). \]

\section{Particle density current for the boosted Maxwell-J\"{u}ttner distribution in the Minkowski spacetime}
\label{appendixc}

In this appendix we derive the counterpart of expressions (\ref{jtabs}, \ref{Jtscat}, \ref{jrjtheta}) for the boosted Maxwell-J\"{u}ttner distribution in the flat Minkowski spacetime expressed in spherical coordinates. They serve as one of the tests of the procedure used in this paper to compute the Vlasov flow in the Schwarzschild spacetime.

We start with the boosted Maxwell-J\"{u}ttner distribution in the flat spacetime in spherical coordinates given by Eq.\ (\ref{boostedf}). It can be written in terms of coordinates $(\tilde \tau,\xi,\theta,\varphi)$ and $(\varepsilon,m,\lambda,\chi)$, as defined in Sec.\ \ref{sec:jcalculation}. This yields
\begin{eqnarray}
 F & = & \alpha \exp\left( - \beta \gamma \varepsilon \right) \exp \left( - \beta \gamma v \epsilon_r \sqrt{\varepsilon^2 - 1 - \frac{\lambda^2}{\xi^2}} \cos \theta \right) \nonumber \\
 && \times \exp \left( \beta \gamma v \frac{\sin \theta}{\xi} \lambda \cos \chi \right).
 \label{fmink}
 \end{eqnarray}
 
 There is no division into absorbed and scattered trajectories in the Minkowski spacetime---all trajectories are simply straight lines. Since the dimensionless radial effective potential reads simply
 \[ U_\lambda(\xi) = 1 + \frac{\lambda^2}{\xi^2},  \]
 the motion is possible for particles with $\varepsilon \ge 1$ with $0 \le \lambda < \lambda_\mathrm{max}(\xi,\varepsilon)$, where
 \[ \lambda_\mathrm{max}(\xi,\varepsilon) = \xi \sqrt{\varepsilon^2 - 1}. \]

A calculation similar to the one described in Sec.\ \ref{sec:jcalculation} yields now
\begin{widetext}
\begin{eqnarray*}
J_t & = & - \frac{4 \pi \alpha m^4}{\xi^2} \int_1^\infty d \varepsilon \varepsilon e^{- \beta \gamma \varepsilon} \int_0^{\lambda_\mathrm{max}} d \lambda \frac{\lambda}{\sqrt{\varepsilon^2 - 1 - \frac{\lambda^2}{\xi^2}}} \cosh \left( \beta \gamma v \sqrt{\varepsilon^2 - 1 - \frac{\lambda^2}{\xi^2}} \cos \theta \right) I_0 \left( \beta \gamma v \lambda \frac{\sin \theta}{\xi} \right),\\
J_r & = & - \frac{4 \pi \alpha m^4}{\xi^2} \int_1^\infty d \varepsilon e^{- \beta \gamma \varepsilon} \int_0^{\lambda_\mathrm{max}} d \lambda \lambda \sinh \left( \beta \gamma v \sqrt{\varepsilon^2 - 1 - \frac{\lambda^2}{\xi^2}} \cos \theta \right) I_0 \left( \beta \gamma v \lambda \frac{\sin \theta}{\xi} \right),\\
J_\theta & = & \frac{4 \pi \alpha m^4 M}{\xi^2} \int_1^\infty d \varepsilon e^{- \beta \gamma \varepsilon} \int_0^{\lambda_\mathrm{max}} d \lambda \frac{\lambda^2}{\sqrt{\varepsilon^2 - 1 - \frac{\lambda^2}{\xi^2}}} \cosh \left( \beta \gamma v \sqrt{\varepsilon^2 - 1 - \frac{\lambda^2}{\xi^2}} \cos \theta \right) I_1 \left( \beta \gamma v \lambda \frac{\sin \theta}{\xi} \right).
\end{eqnarray*}
\end{widetext}
The parameter $M$ appearing in the expression for $J_\theta$ can be any positive constant such that $r = M \xi$, and $\xi$ is dimensionless.

The particle number density reads
\[ n = \sqrt{(J_t)^2 - (J_r)^2 - \frac{(J_\theta)^2}{M^2 \xi^2}}. \]
It can be checked numerically that it is a constant value (independent of $\xi$ and $\theta$), given exactly by Eq.\ (\ref{ninfty}).

Another test is to start with the asymptotic expression for the distribution function (\ref{factionangle}), written in terms of the action-angle variables $(Q^\mu,P_\nu)$, and substitute the expressions for $(Q^\mu,P_\nu)$ given by Eqs.\ (\ref{pmink}) and (\ref{qmink}). By a lengthy but straightforward calculation, one can show that this leads precisely to expression (\ref{fmink}).

\end{document}